\documentclass{article} 
\usepackage{iclr2027_conference,times}

\usepackage{amsmath,amsfonts,bm}

\def\eqref#1{equation~\ref{#1}}

\def\1{\bm{1}}

\DeclareMathAlphabet{\mathsfit}{\encodingdefault}{\sfdefault}{m}{sl}
\SetMathAlphabet{\mathsfit}{bold}{\encodingdefault}{\sfdefault}{bx}{n}

\newcommand{\R}{\mathbb{R}}

\usepackage{hyperref}
\usepackage{url}
\usepackage{graphicx}
\usepackage{siunitx} 

\usepackage{cleveref}
\usepackage{enumitem}

\usepackage{bm}

\newcommand{\Lx}{L^{x}}
\newcommand{\Lout}{L^{x,\text{out}}}
\newcommand{\Lin}{L^{x,\text{in}}}
\newcommand{\LinT}{(L^{\text{in}})^{\top}}
\newcommand{\Lio}{L^{x,\text{io}}}
\newcommand{\Lu}{L^{u}}
\newcommand{\Luout}{L^{u,\text{out}}}
\newcommand{\Luin}{L^{u,\text{in}}}
\newcommand{\Luio}{L^{u,\text{io}}}
\newcommand{\ball}{\text{B}}
\newcommand{\rect}{\text{Rect}}

\newcommand{\cfset}{\mathcal{U}}

\newcommand{\norm}[1]{\left\lVert #1\right\rVert}

\newcommand{\dataset}{\mathcal{D}}

\newcommand{\xhdr}[1]{{\noindent\bfseries #1}.}

\newcommand{\usearch}{\mathcal{U}_{\text{search}}}
\newcommand{\originalDDH}{\text{state-only DDH }}
\newcommand{\tightenedDDH}{\text{state-control DDH }}

\newcommand{\topkDDH}{\text{k-tightened DDH }}

\newcommand{\loss}{\mathcal{L}}
\newcommand{\losspde}{\mathcal{L}_{\text{PDE}}}
\newcommand{\lossterminal}{\mathcal{L}_{\text{terminal}}}
\newcommand{\lossrollout}{\mathcal{L}_{\text{rollout}}}

\newcommand{\xdim}{n_x}
\newcommand{\udim}{n_u}

\newcommand{\datasetPDE}{\mathcal{D}_{\text{PDE}}}
\newcommand{\datasetSUP}{\mathcal{D}_{\text{sup}}}
\newcommand{\datasetTERM}{\mathcal{D}_{\text{term}}}

\newcommand{\uncertainset}{\mathcal{E}}

\DeclareMathOperator*{\argmax}{arg\,max}
\DeclareMathOperator*{\argmin}{arg\,min}

\usepackage{vector}
\usepackage{booktabs}
\usepackage{placeins}
\usepackage{algorithm}
\usepackage{algpseudocode}
\usepackage{tikz}

\usepackage{subcaption}
\title{Data-Driven Hamiltonian \\ for
Neural Optimal Control}

\author{Christopher A. Strong \\
University of California, Berkeley\\
\texttt{christopher\_strong@berkeley.edu}
\And
Aswin Surya \\
University of California, Berkeley\\
\texttt{aswinsurya@berkeley.edu}
\AND
Namhoon Cho\textsuperscript{*} \\
Seoul National University \\
\texttt{namhoon.cho@snu.ac.kr} \\
\And
Jason J. Choi\textsuperscript{*} \\
University of California, Los Angeles \\
\texttt{jjhchoi@ucla.edu} \\
\And 
Claire J. Tomlin\textsuperscript{*} \\ 
University of California, Berkeley \\ 
\texttt{tomlin@berkeley.edu}
}

\iclrfinalcopy 
\begin{document}

\maketitle
{\renewcommand{\thefootnote}{*}\footnotetext{Equal advising.}}

\begin{abstract}

Reinforcement learning (RL) has emerged as a prominent data-driven approach to optimal control, but often requires substantial amounts of data and provides limited guarantees when encountering out-of-distribution state-action pairs.
In this paper, we propose an alternative approach grounded in the well-established Hamilton-Jacobi framework in optimal control theory. Rather than requiring an explicit analytical model of the system dynamics to evaluate the Hamiltonian, we replace this with a data-driven Hamiltonian (DDH), which we generalize to a broad class of optimal control problems. 
In contrast to standard RL approaches, the DDH provides a principled mechanism, requiring only minimal prior system information such as a Lipschitz constant, for quantifying and accounting for uncertainty in regions not covered by the data, enabling the derivation of controllers with performance guarantees. Additional prior system knowledge, when available, can be readily incorporated to improve performance. For high-dimensional systems, we further introduce a physics-informed neural approximation of the value function, trained using a novel procedure that generates worst-case fictitious rollouts evaluated through the DDH. On an LQR problem, the DDH reaches near-optimal cost on every dataset we test, including random data on which offline RL methods remain far from optimal, and needs far fewer samples than online RL.
On a nonlinear lunar lander benchmark, the DDH achieves an optimality gap to a trajectory-optimization reference more than $3\times$ smaller than CQL and FQL, and matches the performance of SAC with approximately $37\times$ fewer samples. These results establish the DDH as a promising framework for data-driven optimal control, particularly in settings where available data are limited in quantity or quality.

\end{abstract}

\section{Introduction}
While optimal control is important across a wide range of applications, including autonomous driving, aeronautics, robotics, HVAC regulation, and chemical process control, the dynamics of such complex systems are often challenging to model accurately.
Reinforcement learning (RL) provides a prominent approach to control under unknown dynamics, as it can learn policies through interactions with the system without requiring an explicit dynamics model. However, RL is often sample inefficient, requiring substantial interaction data to learn an effective policy. 
Moreover, learned policies can perform poorly when encountering states and actions that are insufficiently represented in the training distribution, despite extensive efforts to mitigate such distribution shift \citep{fujimoto2019off, kumar2020conservative, yu2020mopo, kang2022lyapunov}. This limitation is particularly relevant in offline RL, where the performance still remains strongly dependent on the quality and coverage of the available data \citep{park2024value}. At the same time, even when constructing an accurate analytical model is difficult, coarse prior information about the system—such as smoothness properties or bounds on the dynamics—is often readily available. Standard model-free RL formulations, however, provide limited mechanisms for directly incorporating such information to improve data-efficiency, out-of-distribution behavior, or performance guarantees.

To address these limitations of existing approaches, we propose an alternative data-driven approach for optimal control problems. Our approach builds on the Hamilton-Jacobi-Bellman (HJB) framework for continuous-time optimal control, where the value function, representing the optimal cost-to-go, is characterized as the solution of the HJB partial differential equation (PDE) \citep{bardi1997optimal}. The optimal policy is then derived from this value function. Standard solution methods for the HJB-PDE require an explicit model of the system dynamics to evaluate the Hamiltonian, whose optimization objective depends directly on the dynamics.
Thus, when the dynamics are unknown, the Hamiltonian cannot be evaluated directly, limiting the applicability of the HJB framework to data-driven control. Previous work addressed this challenge for reachability problems, a particular class of optimal control problems concerned with reaching or avoiding sets of states, by deriving a direct data-driven conservative bound on the Hamiltonian called the Data-Driven Hamiltonian (DDH) \citep{choi2025data}. The DDH quantifies the worst-case uncertainty consistent with the available dataset through a min–max formulation, using only minimal prior system information such as a verified Lipschitz constant of the vector field. Through this formulation, DDH values at unseen state-action pairs provide conservative bounds on the unknown true  Hamiltonian. This conservative Hamiltonian can then be incorporated into the HJB-PDE, producing an estimated safe set from which the reachability task is guaranteed to be achievable.

We first extend the DDH to a broader class of optimal control problems by incorporating \textit{running costs}, which are common in RL problems but absent from reachability problems. To scale our approach to high-dimensional systems, we further incorporate the DDH into the learning of neural value functions that approximate solutions to the HJB-PDE. Crucially, the PDE residual, which serves as the primary training signal for these value functions \citep{karniadakis2021physics}, can be evaluated at arbitrary states using the DDH. This enables dense supervision throughout the entire state space, including regions not covered by the available data, and thereby facilitates learning a reliable value function. This stands in sharp contrast to standard model-free deep RL, where Bellman-based training signals are evaluated only on sampled transitions, and are thus fundamentally tied to states and actions represented in the collected experience or offline dataset. Similarly, rollouts in model-based RL methods \citep{janner2019trust, yu2020mopo}, which use learned dynamics models, are typically initialized from observed states and kept short to limit the effect of out-of-distribution model error.

Next, to further improve the accuracy of the learned solution, we propose a novel algorithm that augments the PDE residual loss with an additional supervised training signal, analogous to Monte Carlo targets commonly used in RL to improve the reliability of value function learning \citep{sutton2018reinforcement}. By exploiting the min–max formulation underlying the DDH, we generate long-horizon fictitious rollouts that capture the effect of worst-case dynamics uncertainty consistent with the available data, and use the resulting trajectories to construct empirical value targets for direct supervision. This idea is inspired by MPC-guided value-function training in \citet{feng2025bridging}; crucially, however, our fictitious rollouts require neither an explicit dynamics model nor access to a simulator and can be initialized from arbitrary states in the state space.

We validate our approach on a two-dimensional LQR problem and compare the DDH's performance to a variety of RL algorithms.
We find that the DDH performs robustly across datasets of different qualities, especially outperforming the offline RL algorithms on suboptimal demonstrations. This supports our intuition that the DDH is well-suited to tasks where the state-action distribution in the dataset differs substantially from the one the optimal policy visits. With a dataset consisting of random trajectories, the DDH is within 0.1\% of the optimal cost, whereas the offline RL methods CQL \citep{kumar2020conservative}, IQL \citep{kostrikov2021offline}, and FQL \citep{park2025flow} remain $13\%$, $26\%$, and $61\%$ worse than optimal. 

Against online RL, SAC \citep{haarnoja2018soft2} and PPO \citep{schulman2017proximal}, the DDH matches their best performance on every dataset with at least $25\times$ and $100\times$ fewer samples, respectively.
We also evaluate the neural DDH on a 6-D lunar lander system. We find that the neural DDH reduces the optimality gap to a trajectory-optimization reference by more than $3\times$ compared to CQL and FQL, and matches the performance of SAC with approximately $37\times$ and PPO with $2{,}300\times$ less data.

\textit{Related Work.} The most relevant related work is referenced above, and a detailed overview of the related areas is given in \Cref{sec:related_work}.

\textit{Notations.}
$\norm{\cdot}$ is the $l_2$ norm and $|\cdot|$ denotes an absolute value. $\ball(\vect{c}, r) = \{\vect{v} \mid \norm{\vect{v} - \vect{c}} \le r \}$ denotes an $l_2$ ball centered at $\vect{c}$ with radius $r$. $\rect(\vect{c}, \vect{r}) \!=\!\{\vect{v} \mid |v_j - c_j| \le r_j, \forall j\!=\!1, \cdots, n \}$ denotes a hyperrectangle centered at $\vect{c}$ with radii $\vect{r}\!\in\!\R^{n}$. The superscript $i$ denotes the index of a datapoint and the subscript $j$ denotes the $j$-th element of a vector. Vectors are written in bold.

\section{Problem Formulation and Background}
\label{sec:problem formulation}

\xhdr{System Definition and Assumptions}
We consider nonlinear system dynamics 
\[
    \dot{\vect{x}}(t) = f(\vect{x}(t), \vect{u}(t)), \;\; \vect{x}(t_0) = \vect{x}_0
\]
with state $\vect{x}(t) \in \R^{\xdim}$, control $\vect{u}(t) \in \cfset(\vect{x}(t)) \subseteq \R^{\udim}$, and initial state $\vect{x}_0$ at time $t_0$, and where $\cfset(\vect{x})$ is a state-dependent set of admissible controls. The vector field $f$ is assumed to be Lipschitz continuous
in the state. 
While the dynamics $f$ itself is deemed unknown, we assume that (i) a dataset of trajectories from this uncertain system is given as
    $\dataset=\{(\vect{x}^{(i)}, \vect{u}^{(i)}, \vect{v}^{(i)})\}_{i=1}^N,$ 
where $\vect{v}^{(i)} := f(\vect{x}^{(i)}, \vect{u}^{(i)})$ denotes ``state velocity'' and is known exactly, and (ii) we have access to some high-level information about the system that allows us to bound the uncertainty on the dynamics at a given state, for example the state Lipschitz constant $\Lx$ of $f$:
$
    \norm{f(\vect{x}, \vect{u}) - f(\vect{x}', \vect{u})} \le \Lx \norm{\vect{x} - \vect{x}'}.
    \label{eq:lipschitz-uniform}
$
We also define the control Lipschitz constant $\Lu$ of $f$ as $\norm{f(\vect{x}, \vect{u}) - f(\vect{x}, \vect{u}')} \le \Lu \norm{\vect{u} - \vect{u}'}$.
Knowing or estimating such constants requires far less modeling effort than identifying $f$, although it makes our method not fully model-free. We will use $\Lx$ and $\Lu$ as our running example throughout this paper of a known property of the system that allows us to bound the uncertainty, but we provide several other examples in \Cref{appendix:other-uncertainty-sets}.

\xhdr{Optimal Control Problem}
We consider a continuous-time optimal control problem with a running cost. We start at time $t_0$ and end at time $t_f$, where $t_f \in \R \cup \{\infty\}$. $\cfset_{t_0}^{t_f}$ denotes the set of control signals from time $t_0$ to $t_f$. The cost function we optimize over is
\begin{equation}
     J(\vect{x}_0, \vect{u}(\cdot); t_0, t_f) = \int_{t_0}^{t_f} e^{-\gamma (t-t_0)}  c(\vect{x}(t), \vect{u}(t)) dt + e^{-\gamma (t_f-t_0)}c_f\left(\vect{x}(t_f)\right)
\end{equation}
with discount factor $\gamma \ge 0$. We will refer to the function $c$ as the running cost and $c_f$ as the terminal cost. 
Our optimal control problem starting from state $\vect{x}_0$ at time $t_0$ is written as 
$
\label{eq:optimalcontrolproblem}
    \inf_{\vect{u}(\cdot) \in \cfset_{t_0}^{t_f}} J(\vect{x}_0, \vect{u}(\cdot); t_0, t_f).
$ Our goal is to approximately solve this problem given access to our dataset $\dataset$ of state transitions. 

\xhdr{Hamilton-Jacobi-Bellman Equation} The Hamilton-Jacobi-Bellman (HJB) framework solves optimal control problems through dynamic programming  \citep{evans2022partial}. We first define our value function $V(\vect{x}, t)$ as the optimal cost starting from state $\vect{x}$ at time $t$: 
$
    V(\vect{x}, t) = \inf_{\vect{u} \in \cfset_{t}^{t_f}} J(\vect{x}, \vect{u}(\cdot); t, t_f).
$
Applying the principle of dynamic programming yields the HJB-PDE with terminal condition
\begin{equation}
\label{eq:hjpde}
    \frac{\partial V}{\partial t}(\vect{x}, t) = -\left(H(\vect{x}, \vect{p}) - \gamma V(\vect{x}, t)\right); \;\;\;\; V(\vect{x}, t_f) = c_f(\vect{x}),
\end{equation}
where $\vect{p} = \frac{\partial}{\partial \vect{x}} V(\vect{x}, t)$ and the Hamiltonian $H(\vect{x}, \vect{p})$ is
\begin{equation}
\label{eq:hamiltonian}
    H(\vect{x}, \vect{p}) = \min_{\vect{u} \in \cfset(\vect{x})} \vect{p}^\top f(\vect{x}, \vect{u}) + c(\vect{x}, \vect{u}).
\end{equation}
The optimal control at state $\vect{x}$ is given by 
$\vect{u}^* = \argmin_{\vect{u} \in \cfset(\vect{x})} \vect{p}^\top f(\vect{x}, \vect{u}) + c(\vect{x}, \vect{u})$.
More formally, the value function is the unique viscosity solution to the HJB-PDE \citep{bardi1997optimal}\footnote{When we refer to a potentially non-differentiable function $V$ as a solution to a PDE, we interpret it as a weak solution. This notion is formalized through the theory of viscosity solutions \citep{bardi1997optimal}.}. 
For an infinite-time problem with discounting and no terminal cost, the static value function arrived at as $t_f \to \infty$ is uniquely defined by the static discounted HJB-PDE:
$
    H(\vect{x}, \vect{p}) - \gamma V(\vect{x}) = 0.
$

To solve \Cref{eq:hamiltonian} directly, we require knowledge of the dynamics $f$. 
As a result, the Hamiltonian is the key entity we will need to replace when our dynamics are unknown.

\xhdr{Data-Driven Hamiltonian}
The previous work of \citet{choi2025data} focused on bounding the Hamiltonian for reachability problems, which concern driving a system to reach or avoid particular sets of states and have zero running cost. It introduced the Data-Driven Hamiltonian, which provides a conservative bound on the Hamiltonian in the form of 
\begin{equation}
\label{eq:original-ddh-choi}
    \min_{i \in \{1, \hdots, N\}} \max_{\vect{v} \in \uncertainset(\vect{x}; \vect{x}^{(i)})} \vect{p}^\top \vect{v},
\end{equation}
where $\uncertainset(\vect{x}; \vect{x}^{(i)})$ must include the true state velocity at state $\vect{x}$ under control $\vect{u}^{(i)}$. Our approach generalizes this definition of the Data-Driven Hamiltonian and uncertainty set. 

\section{Approach: The Data-Driven Hamiltonian for Optimal Control}

Extending \citet{choi2025data}, we approach our problem by upper-bounding our Hamiltonian. This conservative estimate of the Hamiltonian results in a conservative value function that upper-bounds the true value function. We will loosely refer to any function $\widehat{H}(\vect{x}, \vect{p})$ that provides an upper bound on the Hamiltonian from the data, without requiring the exact system dynamics, as a data-driven Hamiltonian (DDH).

We introduce one form that such an upper bound can take. We start with the concept of an \textit{uncertainty set} $\uncertainset(\vect{x}, \vect{u}; \dataset)$. This represents a set of possible state velocities that could occur at state $\vect{x}$ with control $\vect{u}$. The only requirement of this set is that it must include the true state velocity $f(\vect{x}, \vect{u}) \in \uncertainset(\vect{x}, \vect{u}; \dataset)$.
For instance, given a Lipschitz constant on the state $\Lx$, control $\Lu$, and a single datapoint $(\vect{x}^0, \vect{u}^0, \vect{v}^0)$ we could define 
$
    \uncertainset(\vect{x}, \vect{u}; \dataset) = \ball(\vect{v}^0, \Lx \norm{\vect{x} - \vect{x}^0} + \Lu \norm{\vect{u} - \vect{u}^0}).
$
We also introduce a search set $\usearch(\vect{x}) \subseteq \cfset(\vect{x})$, which represents the set of controls under consideration. We can then define a generalization of \Cref{eq:original-ddh-choi} \citep{choi2025data}:
\par\vspace{6pt}
\hrule
\vspace{1pt}
\noindent\textbf{Data-Driven Hamiltonian (DDH):}\hfill
$\displaystyle \widehat{H}(\vect{x}, \vect{p}) := \min_{\vect{u} \in \usearch(\vect{x})} \; \max_{\vect{v} \in \uncertainset(\vect{x}, \vect{u}; \dataset)} \vect{p}^\top \vect{v} + c(\vect{x}, \vect{u})$%
\hfill\refstepcounter{equation}(\theequation)\label{eq:ddh_general}
\par\vspace{2pt}
\hrule
\vspace{-2pt}
\noindent
The corresponding control will be the optimal $\vect{u}$ for this problem.
Any uncertainty set from \citet{choi2025data} that bounds $\vect{p}^\top \vect{v}$ can be applied here
since $c(\vect{x}, \vect{u})$ does not depend on $\vect{v}$.
We see that the DDH upper-bounds the Hamiltonian since
\[
     \widehat{H}(\vect{x}, \vect{p}) \ge \min_{\vect{u} \in \usearch(\vect{x})} \vect{p}^\top f(\vect{x}, \vect{u}) + c(\vect{x}, \vect{u}) \ge \min_{\vect{u} \in \cfset} \vect{p}^\top f(\vect{x}, \vect{u}) + c(\vect{x}, \vect{u}) = H(\vect{x}, \vect{p}),
\]
where the first inequality holds since $f(\vect{x}, \vect{u}) \in \uncertainset(\vect{x}, \vect{u}; \dataset)$, and the second holds since $\usearch(\vect{x}) \subseteq \cfset$.
If we use the DDH in place of the Hamiltonian when solving for a value function,
the resulting value function is guaranteed to be a conservative bound on the true value function, $\widehat{V}(\vect{x}, t) \ge V(\vect{x}, t)$, following a similar argument to that in \citet{choi2025data}. 
We will instantiate several example uncertainty and search sets in the case where we know $\Lx$ and $\Lu$, as defined in \Cref{sec:problem formulation}. For examples with other system information, see \Cref{appendix:other-uncertainty-sets}.

\xhdr{State-Only} To recover a similar DDH to that in \citet{choi2025data} we can define a DDH from the $i$th datapoint as $\widehat{H}_i$ with
$
    \usearch = \{\vect{u}^{(i)} \} $
    and $\uncertainset^{(i)}(\vect{x}, \vect{u}^{(i)}; \dataset) = \ball\left(\vect{v}^{(i)}, \Lx \norm{\vect{x} - \vect{x}^{(i)}}\right) $
and then take the minimum over the bound from each datapoint as in \Cref{eq:ddh-composition} (see \Cref{appendix:modular-ddhs}).
We will refer to this as the \textit{{\originalDDH}}.
This results in a computationally tractable DDH since $\widehat{H}_i$  maximizes a linear function over a ball, which has a closed-form solution.

\xhdr{State-Control} One limitation of the \originalDDH is that the control obtained from the DDH must be a control already observed in the dataset from the definition of its search set. This can lead to a chattering control signal and jittery movement. We can search outside these observed controls and provide a tighter bound by introducing the following \textit{\tightenedDDH}, with $\widehat{H}_i$ defined by 
$
    \usearch^{(i)} = \cfset $ and $
    \uncertainset^{(i)}(\vect{x}, \vect{u}; \dataset) =
    \ball\left(\vect{v}^{(i)}, \Lx \norm{\vect{x} - \vect{x}^{(i)}} + \Lu \norm{\vect{u} - \vect{u}^{(i)}}\right)
$.
Here we now search over the whole control space, making use of $\Lu$ to bound the behavior of the system on arbitrary controls.  
This is a strict tightening on the \originalDDH, since we have the same uncertainty set but search over a larger search set. If each $\widehat{H}_i$ is expensive to compute, we can choose to only apply this tightening to a subset of the datapoints. One heuristic to choose these datapoints is via the datapoints that achieve the top $k$ bounds in the \originalDDH. We will refer to the DDH where we first compute the \originalDDH then tighten only its top $k$ bounds as the \textit{\topkDDH}.

\subsection{Neural Data-Driven Hamiltonian-Based Value Function}
Grid-based solvers for the HJB-PDE scale exponentially with the state dimension, so for high-dimensional systems, we use a neural network to approximate the solution of the PDE. This resembles minimizing a Bellman residual, except that the residual comes from differentiating the network rather than from a sampled next state. Because the DDH is defined at every state, this residual can be evaluated anywhere in the state space, not only at states in the dataset, which is a key advantage over most model-free RL methods.
We start by introducing a parameterized approximation of the value function, the solution to the HJB-PDE, $V_\theta(\vect{x}, t)$. Since we want the learned value function to be a solution to the HJB-PDE, it needs to satisfy both the terminal condition and PDE given in \Cref{eq:hjpde}.
This motivates two components of the loss function: a PDE loss and a terminal condition loss inspired by \citet{bansal2021deepreach}:
\[
    \losspde(\vect{x}, t) = \left| \frac{\partial V}{\partial t}(\vect{x}, t) + \widehat{H}(\vect{x}, \vect{p}) - \gamma V(\vect{x}, t) \right|, \qquad     \lossterminal(\vect{x}) = \left| V(\vect{x}, t_f) - c_f(\vect{x}) \right|  
\]

The PDE loss enforces only local consistency; thus, as in one-step temporal-difference learning, errors can propagate slowly. \citet{feng2025bridging} addressed this in reachability by supervising the network with returns from MPC rollouts, which require a dynamics model. We instead propose an approach to generate fictitious rollouts using the worst-case dynamics implicit in our DDH.
We will work with the form of the DDH in \Cref{eq:ddh_general}.
Let 
\[
    \vect{u}^*(\vect{x}, \vect{p}), \vect{v}^*(\vect{x}, \vect{p}) = \argmin_{\vect{u} \in \usearch(\vect{x})} \argmax_{\vect{v} \in \uncertainset(\vect{x}, \vect{u}; \dataset)} \vect{p}^\top \vect{v} + c(\vect{x}, \vect{u}).
\] 
We can then define adversarial dynamics that depend only on the dataset and $V_\theta$:
$
    f_{\text{adversarial}}(\vect{x}, \vect{u}^*(\vect{x}, \vect{p})) = \vect{v}^*(\vect{x}, \vect{p}),
$
where $\vect{p} = \nabla_{\vect{x}} V_\theta(\vect{x}, t)$. This corresponds to the fictitious dynamics introduced in the proof of Theorem 1 in \citet{choi2025data}, resulting from a differential game where the leader chooses the control $\vect{u}^*$ and an adversarial follower chooses the worst-case realization of the dynamics $\vect{v}^*$ subject to the uncertainty set resulting from the choice of $\vect{u}^*$.
We can then perform rollouts of the system using these adversarial dynamics despite not knowing the dynamics of the true system.
Let $v_{\text{sup}}(\vect{x}, t)$ be the discounted accumulated cost obtained from a rollout starting at state $\vect{x}$ at time $t$ under the adversarial dynamics, plus the discounted value from $V_\theta$ at the rollout's final state. For $V_\theta$ to be consistent we would like $V_\theta(\vect{x}, t) = v_{\text{sup}}(\vect{x}, t)$. Formally, since $f_{\text{adversarial}}$ corresponds to the adversarial dynamics from \citet{choi2025data}, if $V_\theta$ perfectly represented the desired $\widehat{V}$ then $v_{\text{sup}}(\vect{x}, t) = \widehat{V}(\vect{x}, t) \ge V(\vect{x}, t)$.
We define the supervision loss as 
$
    \lossrollout(\vect{x}, t) = \left|V_\theta(\vect{x}, t) - v_{\text{sup}}(\vect{x}, t) \right|.
$
This can be thought of as policy evaluation in a policy-iteration algorithm, as it evaluates the current policy and pulls the value function towards that value.

During training we sample datasets of state-time pairs for each of our losses, which we will denote as $\datasetPDE$, $\datasetTERM$, and $\datasetSUP$. As in \citet{bansal2021deepreach}, a curriculum learning approach can be applied to the choice of $t$ in the samples.
Altogether, our empirical loss is
\[
    \loss(\theta) = \frac{1}{|\datasetPDE|}\sum_{\vect{x}_i, t_i \in \datasetPDE} \losspde(\vect{x}, t) + \frac{\lambda_1}{|\datasetTERM|} \sum_{\vect{x}_i \in \datasetTERM} \lossterminal(\vect{x})+  \frac{\lambda_2}{|\datasetSUP|} \sum_{\vect{x}_i, t_i \in \datasetSUP} \lossrollout(\vect{x}, t)
\]
with $\lambda_1, \lambda_2$ weights we can set to prioritize different losses. Since the rollouts are comparatively expensive to compute, the dataset $\datasetSUP$ can be fixed for a number of iterations before being refreshed.

\begin{algorithm}[t]
\caption{Neural DDH training (static, discounted)}
\label{alg:neural-ddh}
\begin{algorithmic}[1]
\Require dataset $\dataset$, sensitivity bounds, running cost $c$, discount rate $\gamma$, weight $\lambda_2$, refresh period $T_{\text{sup}}$, rollout horizon $\tau$, step $\Delta t$
\State initialize $\theta$; $\datasetSUP \gets \emptyset$
\For{$j = 1, 2, \hdots$}
    \State sample $\datasetPDE$ uniformly in $\mathcal{X}$
    \If{$j \bmod T_{\text{sup}} = 0$}
        \State sample start states $\vect{x}_0$ in a supervision region $\mathcal{X}_{\text{sup}} \subseteq \mathcal{X}$
        \State roll out for $\tau / \Delta t$ steps, stopping early if the next state leaves $\mathcal{X}$: set $\vect{p} = \nabla_{\vect{x}} V_\theta(\vect{x})$, take $(\vect{u}^*, \vect{v}^*)$ from \Cref{eq:ddh_general}, step $\vect{x} \gets \vect{x} + \vect{v}^* \Delta t$
        \State $\datasetSUP \gets \{(\vect{x}_0, v_{\text{sup}})\}$, with $ v_{\text{sup}} = \sum_{k<K} e^{-\gamma k \Delta t} c(\vect{x}_k, \vect{u}^*_k)\Delta t +  e^{-\gamma K \Delta t} V_\theta(\vect{x}_K)$,  where $\vect{x}_K$ is the final state (a fixed exit value is used if the rollout leaves $\mathcal{X}$)
    \EndIf
    \State $\loss(\theta) = \frac{1}{|\datasetPDE|}\sum_{\datasetPDE} \big|\widehat{H}(\vect{x}, \nabla_{\vect{x}} V_\theta) - \gamma V_\theta(\vect{x})\big| + \frac{\lambda_2}{|\datasetSUP|}\sum_{\datasetSUP} \big|V_\theta(\vect{x}_0) - v_{\text{sup}}\big|$
    \State update $\theta$ with a gradient step on $\loss(\theta)$
\EndFor
\end{algorithmic}
\end{algorithm}
\vspace*{-0.8em}
We can also learn an infinite-horizon static value function $V_\theta(\vect{x})$ when the discount factor $\gamma > 0$. The rollout supervision loss remains unchanged (just losing its dependence on time), while the PDE loss becomes
$
\losspde(\vect{x}) = | \widehat{H}(\vect{x}, \vect{p}) - \gamma V(\vect{x})|
$
encouraging satisfaction of the static HJB equation. In practice, we observe that learning this static value function leads to a more stable training problem, likely due to its contraction property. \Cref{alg:neural-ddh} summarizes the training procedure. Note that our learning process can also be applied to the original reachability DDH in \citet{choi2025data}.

\section{Experimental Results}
\label{sec:experiments}
In this section, we investigate how the DDH compares to existing data-driven optimal control techniques. We begin by validating the DDH on a simple LQR problem. Through this case study, we are able to explore how the \textit{type} of data the DDH has affects its performance, and how this differs from offline RL methods: the DDH appears to benefit from random (less optimal) demonstrations, whereas offline RL algorithms suffer from that randomness.
We also evaluate the neural DDH on a lunar lander problem and see substantial advantages in data-efficiency over model-free RL methods.   

\subsection{Verification on an LQR Problem}
\label{sec:lqr}
We first validate the DDH on a simple two-dimensional linear system with quadratic cost.
The dynamics are $\dot{\vect{x}} = A \vect{x} + B \vect{u}$ with $A = \left[\begin{smallmatrix}0 & 1 \\ 0 & 0\end{smallmatrix}\right]$,
$B = I$, state and control bounds $\mathcal{X} = [-2, 2]^2$ and
$\cfset = [-4, 4]^2$, and finite-horizon LQR cost
$
J(\vect{x}_0, \vect{u}(\cdot)) = \int_0^T (\vect{x}^\top Q \vect{x} + \vect{u}^\top R \vect{u})\, dt
+ \vect{x}(T)^\top Q_f \vect{x}(T)
$
with $Q = R = Q_f = I$, horizon $T = 2\,$s. When simulating the system, we use an Euler approximation with step size $\Delta t = 0.01\,$s. As the optimal reference, we apply the controller $\vect{u} = -K\vect{x}$ with $K$ a constant gain from the finite-horizon Riccati differential equation solution. It achieves $J^\star = 2.96$ on a fixed set of $N_{\text{eval}} = 1{,}000$ initial conditions (ICs).
A random policy attains $\sim 39$ on the same set. A disjoint set of $500$ validation ICs is used for all checkpoint and hyperparameter selection.

\xhdr{DDH Variants}
To isolate the effects of the uncertainty set, tightening, and function approximation, we consider seven DDH variants. Six variants come from the spherical and rectangular uncertainty sets from the $\Lx$ and $\Lio$ Lipschitz constants respectively (see \Cref{appendix:other-uncertainty-sets}) without tightening and with tightening ($k = 1$ or $k = 10$). The final variant uses the neural DDH, with a rectangular uncertainty set from $\Lio$ and tightening of $k=10$. For this system, $\Lio = |A|$ and $\Luio = |B|$ entrywise, so the rectangular set encodes the model up to the sign of each entry, while the spherical set uses only the scalar constants $\Lx = 1$ and, with tightening, $\Lu = 1$.
For this low-dimensional study the neural DDH is trained with the PDE and terminal losses only. 
The full per-variant results are in \Cref{tab:lqr-ddh-variants}, while \Cref{fig:ddh-variants-comparisons} shows the four most informative (both in \Cref{appendix:ddh-variants}).

\xhdr{Datasets}
We train all methods on four datasets (\Cref{fig:lqr-datasets}):
\texttt{optimal}, \texttt{noisy optimal}, \texttt{random trajectories}, and \texttt{uniform random}. For the \texttt{optimal} dataset, the known optimal controller was applied to generate trajectories. For the \texttt{noisy optimal} dataset, the same controller was applied but with additive Gaussian noise. For the \texttt{random trajectories} dataset, a uniformly random control drawn from the control bounds was applied to generate trajectories. For \texttt{uniform random}, states and controls were uniformly randomly sampled, and then their dynamics were evaluated (no trajectories were rolled out). 
Each trajectory dataset contains $100{,}000$ transitions ($500$ trajectories of $200$ steps at $\Delta t = 0.01$), and every result at dataset size $n$ uses the first $n$ transitions, so each larger dataset is a strict superset of the smaller ones.

\xhdr{Baseline Algorithms}
We compare the DDH against conservative Q-learning (CQL) \citep{kumar2020conservative}, implicit Q-learning (IQL) \citep{kostrikov2021offline}, and flow Q-learning (FQL) \citep{park2025flow}, three popular offline RL methods. We chose to compare against offline methods since they most closely match the problem setting we introduce for the DDH, where we have access only to a fixed offline dataset. As a reference for sample efficiency, we also train the online RL methods PPO \citep{schulman2017proximal} and SAC \citep{haarnoja2018soft2} directly on the true system ($\gamma = 0.995$, single seed). Since they collect their own data, they are not a like-for-like comparison. See \Cref{appendix:lqr-baselines} for details on how the baselines were trained.

\begin{figure}[t]
\centering
\includegraphics[width=\linewidth]{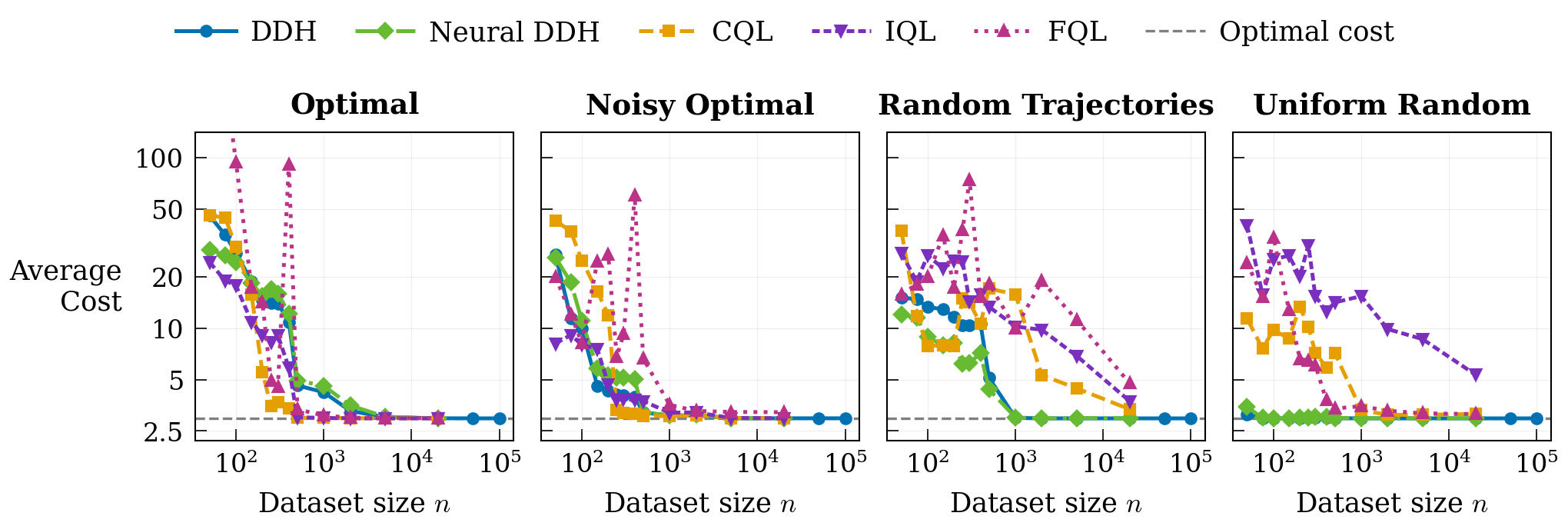}
\caption{\textbf{The DDH and offline RL are data-efficient on opposite datasets.} Average rollout cost over $1{,}000$ test initial conditions versus dataset size $n$ (both log scale); dashed line: optimal reference (from the Riccati equation). FQL's behavior-cloning weight $\alpha$ is selected once per dataset on the validation set at $n = 20{,}000$ and reused for all $n$; two FQL points on \texttt{optimal} exceed the axis. The DDH improves fastest on datasets that cover the state and control space broadly, while offline RL improves fastest on expert data.}
\label{fig:data-efficiency}
\vspace{-0.0em}   
\begin{minipage}[t]{0.55\linewidth}
\vspace{0pt}
\includegraphics[width=\linewidth]{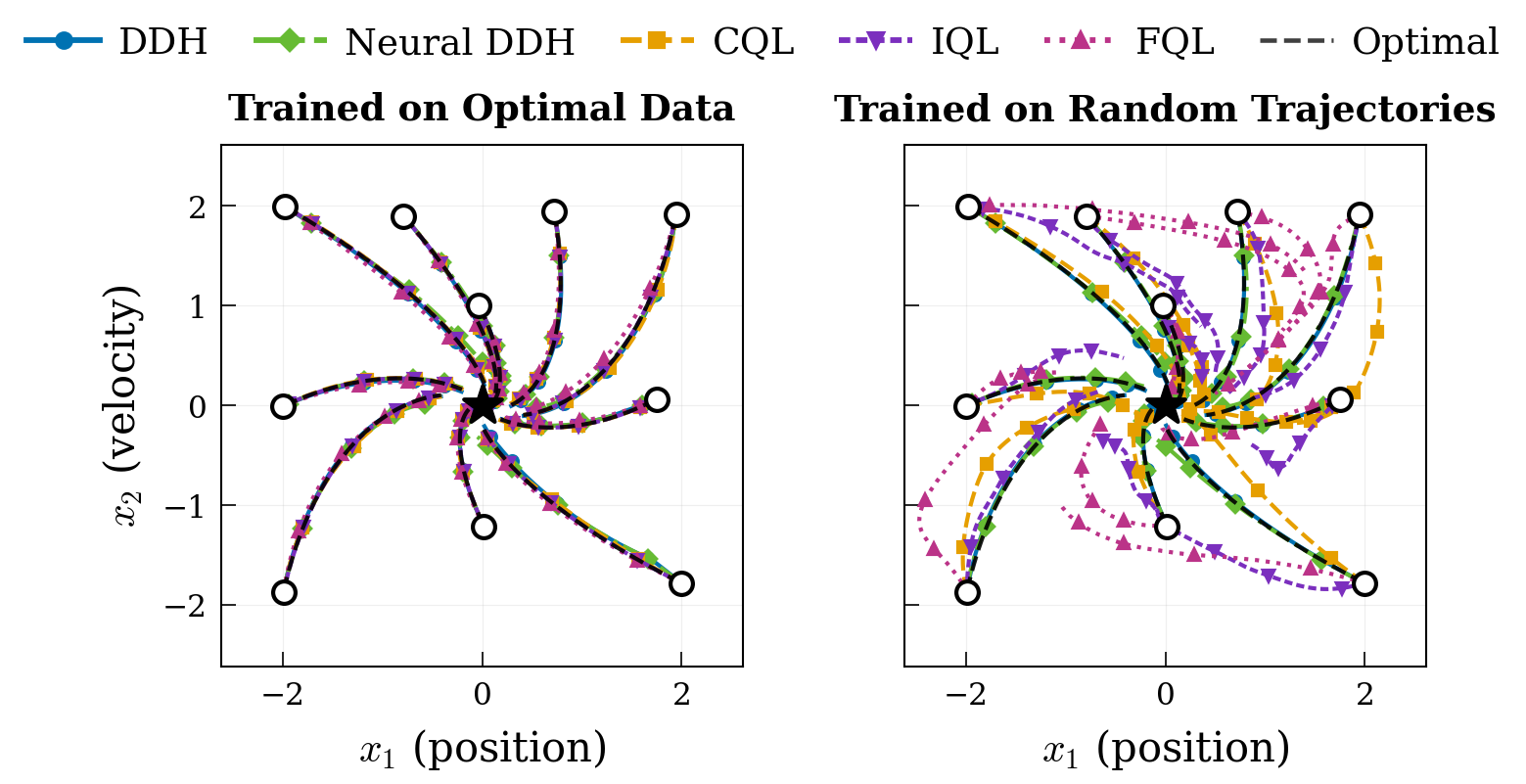}
\end{minipage}\hfill
\begin{minipage}[t]{0.42\linewidth}
\vspace{0pt}
\captionsetup{skip=0pt}
\caption{\textbf{Data quality affects offline RL's behavior but not the DDH's.} Ten trajectories using policies trained on \texttt{optimal} data (left) and \texttt{random trajectories} (right), $n = 20{,}000$. On optimal data all methods perform well. On random data, the DDH matches the optimal trajectory while the other algorithms drift away.}
\label{fig:lqr-trajectories}
\end{minipage}
\vspace{-1.7em}  
\end{figure}

\xhdr{Results} \Cref{fig:data-efficiency} shows each method's cost as the dataset grows. At $n = 20{,}000$ the DDH (rectangular, $k = 10$) is within $0.2\%$ of $J^\star$ on every dataset, while on \texttt{random trajectories} CQL, IQL, and FQL remain $13\%$, $26\%$, and $61\%$ above it. The two families favor opposite datasets. The DDH is within $1.5\%$ of $J^\star$ from $n = 75$ on \texttt{uniform random} (within $4\%$ in each of five dataset re-draws, \Cref{appendix:lqr-baselines}) and within $1\%$ from $n = 1{,}000$ on \texttt{random trajectories}, but needs $n \approx 5{,}000$ to come within $2\%$ on \texttt{optimal}, whose datapoints concentrate along optimal trajectories. CQL is the reverse: within $1\%$ of $J^\star$ by $n = 500$ on \texttt{optimal}, but not converging on \texttt{random trajectories} by $n = 20{,}000$. IQL and FQL show the same pattern (\Cref{fig:data-efficiency,fig:lqr-trajectories}).

One possible explanation lies in how each method uses its data. The DDH uses each datapoint as evidence of how the system responds near that state and control, regardless of whether the recorded action was optimal. Even if an action was suboptimal in that state, it might be useful for neighboring states, and the DDH allows for this extrapolation in a structured way. As a result, while both families benefit from state coverage, the DDH benefits more from coverage of the control space. CQL, IQL, and FQL instead tend to keep their policies closer to the recorded actions (through a value penalty on unseen actions, advantage-weighted regression onto dataset actions, and a behavior-cloning term, respectively), which helps when those actions are optimal and hurts when they are random.

Among the DDH variants at $n = 500$ (\Cref{fig:ddh-variants-comparisons}), the rectangular uncertainty set and tightening each lower the cost on most datasets, and the neural DDH is comparable to the grid solution. Even the spherical set, which uses only scalar Lipschitz constants, outperforms every offline RL baseline on \texttt{random trajectories} from $n = 500$ (\Cref{tab:sphere-curve}). Online RL is far less data-efficient (\Cref{fig:lqr-cost-comparison}): after $50$k (SAC) and $526$k (PPO) environment steps, neither reaches the $2.98$ the DDH attains from $1{,}000$ random transitions.

\subsection{Lunar Lander Case Study}
\label{sec:lunar-lander-results}
We evaluate the neural DDH against a variety of baselines on a 6-D lunar lander system. 
As in \citet{inoue2026lunar}, we model the lunar lander with a 6-D state that includes the lunar lander's 2-D position ($x, y$), its rotation angle ($\theta$), and their corresponding first derivatives ($v_x, v_y, \text{ and } \omega$). The control is 3-D and consists of the main thruster and the right and left side thrusters ($u_1$, $u_2$, and $u_3$). The detailed dynamics, a diagram, and derivations of the associated Lipschitz constants can be found in \Cref{appendix:lunar-lander-details}.
 We solve an infinite-time discounted optimal control problem with $\gamma = 0.33$, known equilibrium thrust $\vect{u}_0$, and running cost
\[
    c(\vect{x}, \vect{u}) = 1 - e^{-\sqrt{\vect{x}^\top Q \vect{x}}} + (\vect{u} - \vect{u}_0)^\top R (\vect{u} - \vect{u}_0), \;\;\;\; Q = 0.01 \cdot \text{diag}(1, 1, 1, 1, 10, 2), \;\;\;\; R = 0.3I.
\]
The state space is $\mathcal{X} = \rect(\vect{0}, [15, 12, 15, 12, \pi, 3\pi])$ over $(x, v_x, y, v_y, \theta, \omega)$, and we evaluate on initial conditions drawn from an inner supervision region $\mathcal{X}_{\text{sup}} = \rect(\vect{0}, [7.5, 6, 7.5, 6, \pi, 7.1])$ from which we also draw the initial conditions for the DDH's supervision loss.

 We apply the neural DDH to this problem with known sensitivity matrix $\Lio$ (see \Cref{appendix:other-uncertainty-sets}) and we further reduce the uncertainty set by incorporating the knowledge that three of the six states are single integrators of the other states (e.g., $\dot{x} = v_x$). Slices of the value function are visualized in \Cref{appendix:learned-value-fcns} and training details are in \Cref{appendix:lander-training}. We compare against several representative offline RL algorithms (CQL and FQL), online RL algorithms (PPO and SAC), and a model-based RL baseline consisting of a system identification step where we fit a neural network model of the dynamics using supervised learning and then a control step where we apply MPC (\Cref{tab:hp-sysid,tab:hp-mpc}). The neural DDH and offline RL algorithms are trained on the same dataset, drawn uniformly at random from the state and control space (like the \texttt{uniform random} dataset we used for the LQR problem). In addition, we implemented a trajectory-optimization baseline that has access to the true dynamics. We treat the performance of the trajectory optimizer as an oracle that we can then compare each method against. We define each method's optimality gap as its average cost minus that of the trajectory optimizer. We evaluate all methods by rolling out $15\,$s long trajectories from a set of $20{,}000$ random initial conditions and compare their average costs. 

\begin{figure}[t]
\centering

\begin{subfigure}[t]{0.44\linewidth}
  \centering
  \caption{2-D LQR problem}
  \label{fig:lqr-cost-comparison}
  \includegraphics[height=2.03in]{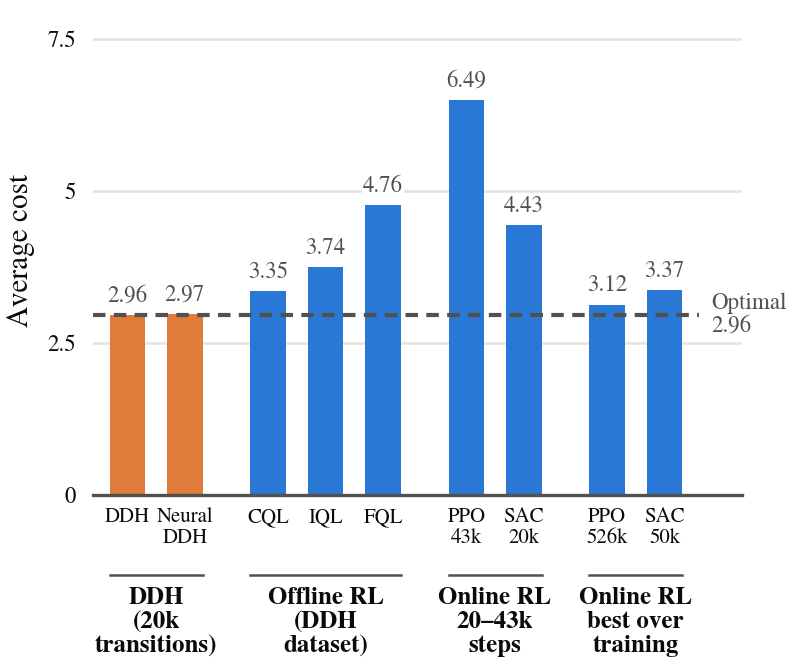}
\end{subfigure}%
\hfill
\begin{subfigure}[t]{0.54\linewidth}
  \centering
  \caption{Lunar lander}
  \label{fig:lunar-lander-cost-comparison}
  \includegraphics[height=2.03in]{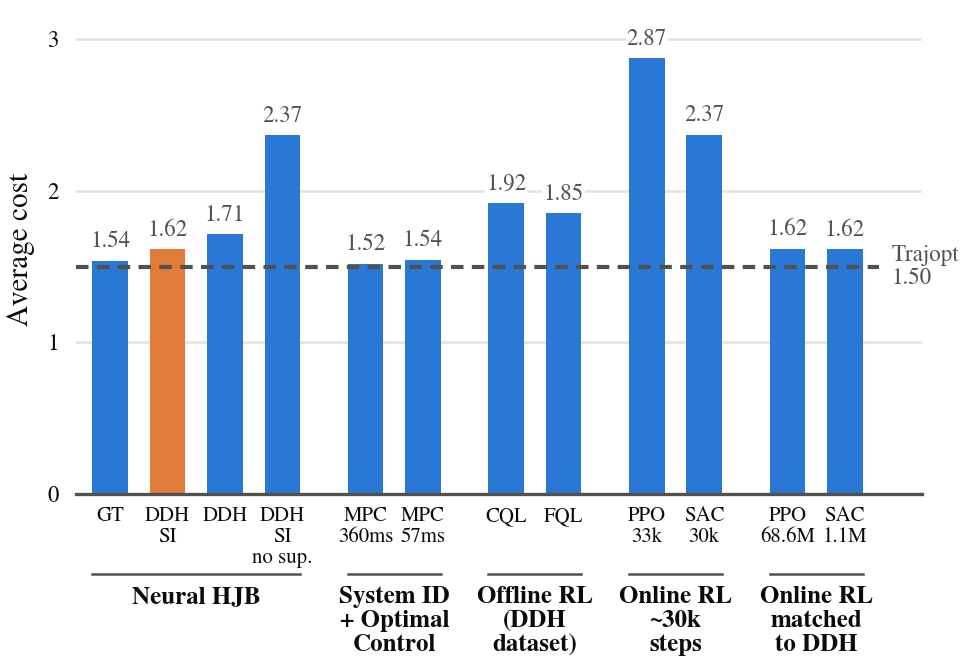}
\end{subfigure}
\caption{\textbf{The DDH outperforms model-free RL.} (a) Average cost on $1$k initial conditions; offline methods use the same $20$k transitions, and online RL interacts with the true system. 
The last group shows PPO and SAC at their best checkpoints, the end of training ($526$k and $50$k steps). Online RL is single-seed; the CQL and FQL gaps hold across three seeds (\Cref{appendix:lqr-baselines}). (b) Average cost on the lunar lander evaluated on $20$k initial conditions from the supervision region. Note that the training dataset was drawn from the full region, of which the supervision region constitutes roughly $4.8\%$. We evaluate on this smaller region so that more rollouts have the potential to be kept within the state space. The dashed line is a trajectory-optimization reference computed with the true dynamics. The DDH has more than a $3\times$ lower optimality gap than the offline RL algorithms, and SAC and PPO need roughly $37\times$ and $2{,}300\times$ more interaction to match the DDH's performance. Fitting a dynamics model and then applying MPC performs better, but must solve a trajectory optimization online at every re-planning step and lacks the built-in guarantees and principled out-of-distribution degradation of the DDH. We show two versions of MPC with different average runtimes.}
\label{fig:cost-comparison}
\vspace{-0.7em}  
\end{figure}
\begin{figure}[t]
\centering
\begin{minipage}[t]{0.47\linewidth}
\vspace{0pt}
\includegraphics[width=0.95\linewidth]{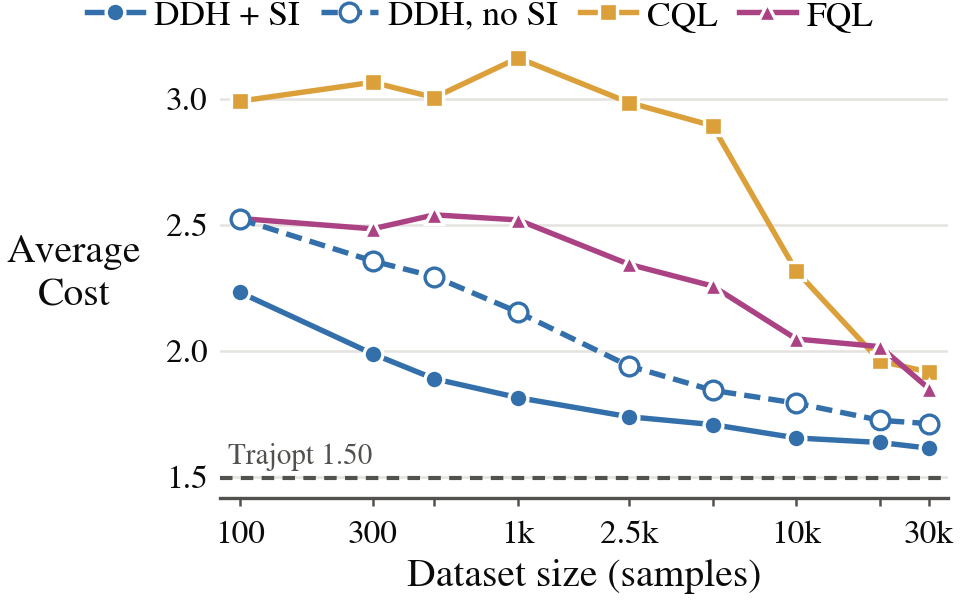}
\end{minipage}\hfill
\begin{minipage}[t]{0.5\linewidth}
\vspace{0pt}
\caption{\textbf{The neural DDH is substantially more data-efficient than both CQL and FQL.} Average cost of the neural DDH, CQL, and FQL as the dataset grows, with and without knowledge of the single-integrator states $x, y, \text{and } \theta$ for the DDH. CQL and FQL show the best cost over $3$ training seeds. With just $1{,}000$ samples, the DDH with known single integrators outperforms CQL and FQL on $30{,}000$.}
\label{fig:lunar-lander-efficiency-fig}
\end{minipage}
\vspace{-1.2em}
\end{figure}
\begin{figure}[t]
\centering
\includegraphics[width=0.9\linewidth]{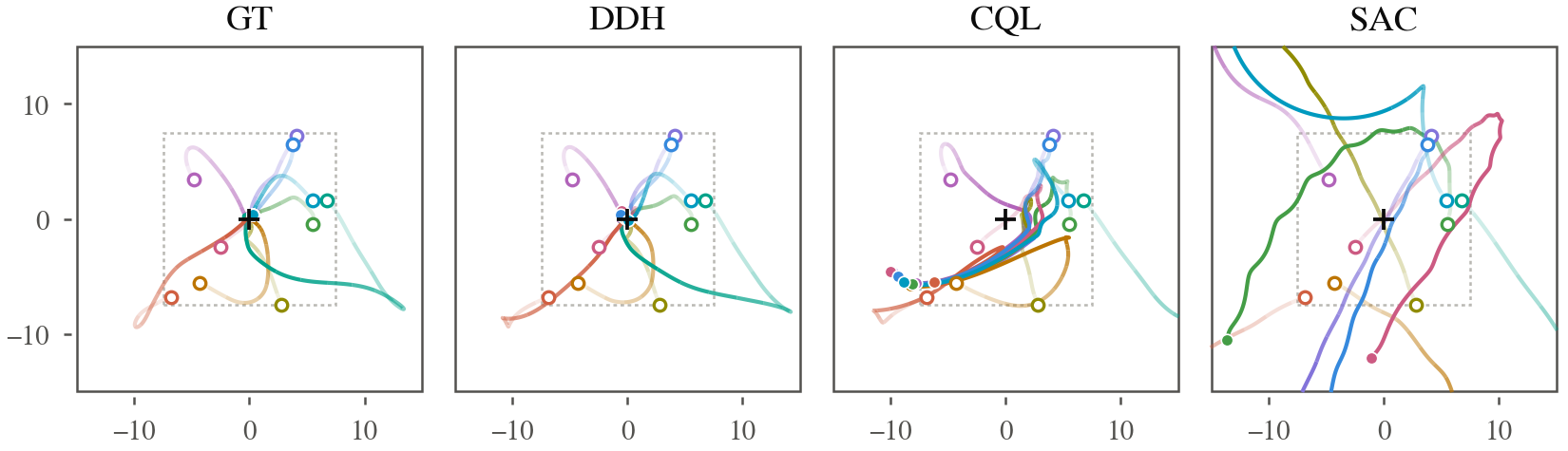}
\caption{\textbf{The DDH's trajectories qualitatively outperform CQL and SAC.} The plots show a set of trajectories rolled out by (i) our neural approach with the true dynamics, (ii) the neural DDH, (iii) CQL, and (iv) SAC. CQL uses the same dataset as the DDH, while SAC uses the same number of samples (30k) but gathered through its online approach. The DDH drives the lander towards the origin, while CQL keeps the lunar lander within bounds but does not drive it to the origin, and SAC fails to keep the lunar lander within bounds.}
\label{fig:lunar-lander-trajectories}
\vspace{-1.0em}  
\end{figure}

\Cref{fig:lunar-lander-cost-comparison} compares the DDH to these baselines. We observe that the DDH substantially outperforms the offline RL algorithms, reducing the optimality gap by $67\%$ compared to FQL and $72\%$ compared to CQL. The model-based baseline, whose learned dynamics model uses no structural prior, outperforms the DDH. It must, however, solve an optimization problem online at every re-planning step, whereas the DDH controller evaluates the value gradient and selects among recorded controls. It also lacks the DDH's guarantee and its well-characterized degradation away from the data.
Online RL given a comparable budget of about $30$k environment steps performs substantially worse; matching the performance of the DDH takes SAC $1.1$M steps and PPO $69$M steps, $37\times$ and $2{,}300\times$ as much data respectively.
\Cref{fig:lunar-lander-cost-comparison} also presents two ablations on the DDH: (i) removing the knowledge of the single integrator states, and (ii) removing the fictitious rollout-based supervision loss. We see that the single integrators clearly improve the DDH, but are less essential than supervision, without which the DDH drastically deteriorates. 
\Cref{fig:lunar-lander-efficiency-fig} breaks down data-efficiency against CQL and FQL: with only $1{,}000$ samples, the DDH outperforms CQL and FQL trained on all $30{,}000$. The trajectories in \Cref{fig:lunar-lander-trajectories} reflect these costs. 


\section{Conclusion and Future Work}
In this work, we extended the data-driven Hamiltonian (DDH) to a broader class of optimal control problems with running costs, and introduced a neural approximation to the solution to the HJB-PDE we call the neural DDH. The neural DDH greatly benefits from a novel approach we introduce to generate worst-case fictitious rollouts for its training. In comparison to most model-free RL algorithms, which only train on states in their dataset, the neural DDH is able to train its network across the entire state space by using the DDH to handle previously unobserved state-action pairs. 

The DDH offers an alternative to the conservative penalties and heuristic uncertainty measures that RL typically applies to unseen states and actions: a worst-case bound computed from known properties of the system. Because this bound is defined everywhere, the value function can be trained everywhere, and prior knowledge can be directly incorporated. We expect this to be useful in robotics and aerospace, wherever some physical structure is known but a full model is not.
We also hypothesize a core difference in what kind of dataset is best for the different methods: the DDH benefits most from broad coverage of states and actions, while offline RL benefits most from optimal demonstrations.
The DDH is therefore a natural fit when consistent and high-quality expert demonstrations are hard to collect. 

The DDH has several limitations. It requires enough system information to create a valid uncertainty set on the dynamics, and a cost function that can be evaluated at any state and action, rather than only on logged transitions. With only minimal information, such as a single Lipschitz constant, the uncertainty set can grow prohibitively large, especially in high-dimensional spaces. Overestimating the bound degrades cost gradually (\Cref{appendix:l-misspec}).
Some directions for future work include extending offline reinforcement learning algorithms (e.g., CQL) to make use of some system information, bounds that combine information across several datapoints, pairing the DDH with a nominal dynamics model with a known maximum error, and extending the approach to accommodate stochastic dynamics.


\subsection*{AI Use Statement}



In this work, we used generative AI tools to support implementing the baseline methods for comparison (via the use of Claude Code during that implementation).
We have not used generative AI tools for generating synthetic data sets, developing theoretical models or conceptual frameworks, formulating mathematical claims, providing critical ingredients for proving mathematical claims, assisting in the writing of proofs, proposing or refining hypotheses, assisting with translation, cleaning and reformatting datasets, supporting qualitative and thematic data analysis, or interpreting results. 

Additionally, we used generative AI tools for creating visualizations while debugging and for creating and editing some miscellaneous areas in our software (again via Claude Code) and writing out some tables in latex. We have also used generative AI tools to cross-check numerical claims against our experimental results and to suggest edits to the abstract, captions, and experimental sections; all such text was reviewed, verified, and finalized by the authors.

We have reviewed all AI-assisted work. We have thoroughly tested our implementation and checked all data used to produce the resulting figures. We take responsibility for the final content of this work,
including text, claims or artifacts produced with the aid of generative AI.


\subsubsection*{Acknowledgments}
We thank Zeyuan Feng for his valuable insights and advice that helped lead us to develop our method for generating fictitious rollouts and Srikar Babu Gadipudi for his patient explanations and conversations around reinforcement learning. This work was supported by the Strategic Hub for International Research Collaboration project of Seoul National University (SNU) and the NSF Safe Learning Enabled Systems Program. 


\bibliography{iclr2027_conference}
\bibliographystyle{iclr2027_conference}

\appendix
\section{Appendix}

\subsection{Related Work}
\label{sec:related_work}

There are a variety of data-driven approaches for controlling systems with unknown dynamics. Reinforcement learning is one of the richest fields concerning the optimal control of unknown systems. In model-free RL, online approaches, such as proximal policy optimization (PPO) \citep{schulman2017proximal} and soft actor-critic (SAC) \citep{haarnoja2018soft2, haarnoja2018soft}, explore the state-action space while learning state or state-action value functions. 
Offline RL approaches rely on a fixed pre-collected dataset of system transitions. Conservative Q-Learning (CQL) \citep{kumar2020conservative} learns a state-action value function while penalizing out-of-distribution (OOD) actions to avoid an overoptimistic value function on OOD data. Implicit Q-Learning (IQL) \citep{kostrikov2021offline} avoids querying OOD actions altogether by fitting the value function with expectile regression over dataset actions. Flow Q-Learning (FQL) \citep{park2025flow}, a more recent approach which learns a state-action value function, a one-step policy, and a flow-matching model of the behavior policy, effectively incorporates information from near-optimal demonstrations from the dataset. However, in all of these approaches, the value function supervision is fundamentally done at observed state transitions, leaving the behavior outside the support of the collected data only indirectly constrained. 

Other offline methods, such as AWR \citep{peng2019advantage} and TD3+BC \citep{fujimoto2021minimalist}, similarly regularize the policy toward the data; we compare against CQL, IQL, and FQL as widely used representatives with strong reported performance.

Model-based RL tackles the same problem, and typically consists of (i) a model-learning stage, where an explicit dynamics model is learned, and (ii) the development of a controller using this model \citep{janner2019trust, chua2018deep}. 
However, model-based RL can also suffer from OOD issues when the dynamics model is used outside of its training data, particularly when restricted to an offline dataset.
Model-Based Offline Policy Optimization (MOPO) 
\citep{yu2020mopo} attempts to mitigate this by penalizing state-action pairs with higher uncertainty. 

There is also a body of work in direct data-driven model predictive control (MPC). However, the majority of this work focuses on linear systems in order to leverage Willems' fundamental lemma \citep{berberich2020data, coulson2019data}, which restricts its generalizability.

The approach that we build on, the Data-Driven Hamiltonian (DDH), proposed by \citet{choi2025data} and extended by \citet{bajelani2026robust}, provides a conservative bound on the Hamiltonian, which allows Hamilton-Jacobi Reachability problems \citep{bansal2017hamilton} to be solved for systems whose dynamics are unknown. We directly extend this approach to the broader class of optimal control problems involving running costs, as well as to higher-dimensional systems, while retaining its key insights that (i) the Hamiltonian is where the system dynamics enter, and (ii) data and minimal system information can be used to bound this Hamiltonian. Finally, recent advances in physics-informed machine learning are applied to solve Hamilton-Jacobi reachability problems \citep{bansal2021deepreach, feng2025bridging, li2025hjrno}, where a model-based Hamiltonian is used to evaluate the PDE residual.

\subsection{Modular Combination of DDHs}
\label{appendix:modular-ddhs}
Given $\{\widehat{H}_i(\vect{x}, \vect{p})\}_{i=1}^n$, where $\widehat{H}_i$ are each valid DDHs, we can define the combined DDH as the minimum of the individual DDHs, since each DDH individually upper-bounds the Hamiltonian:
\begin{equation}
\label{eq:ddh-composition}
   \widehat{H}(\vect{x}, \vect{p}) = \min_i\left\{\widehat{H}_i(\vect{x}, \vect{p}) \right\} \ge H(\vect{x}, \vect{p}).
\end{equation}
For example, if you had a nominal model of your system dynamics $\hat{f}$ with a known maximum error norm $e_{\max}(\vect{x},\vect{u})$, then you could have one DDH based on this nominal model with $\uncertainset(\vect{x}, \vect{u}) = \ball(\hat{f}(\vect{x}, \vect{u}), e_{\max}(\vect{x}, \vect{u}))$
and another from a dataset of observations. 
The combined DDH would then apply the upper bound derived from the dataset when close to the data, and the upper bound from the nominal model when farther from the data.
When combined together in this way, the optimal control and the corresponding worst-case value of $\dot{\vect{x}}$ for the combined Hamiltonian will be the $\argmin$ and $\argmax$ of $\widehat{H}_{i^*}$ where $i^*$ is the index of the minimum $\widehat{H}_i$.

\subsection{Additional Lipschitz Constants and Uncertainty Sets}
\label{appendix:other-uncertainty-sets}

\begin{figure}[t]
\centering
\includegraphics[width=\linewidth]{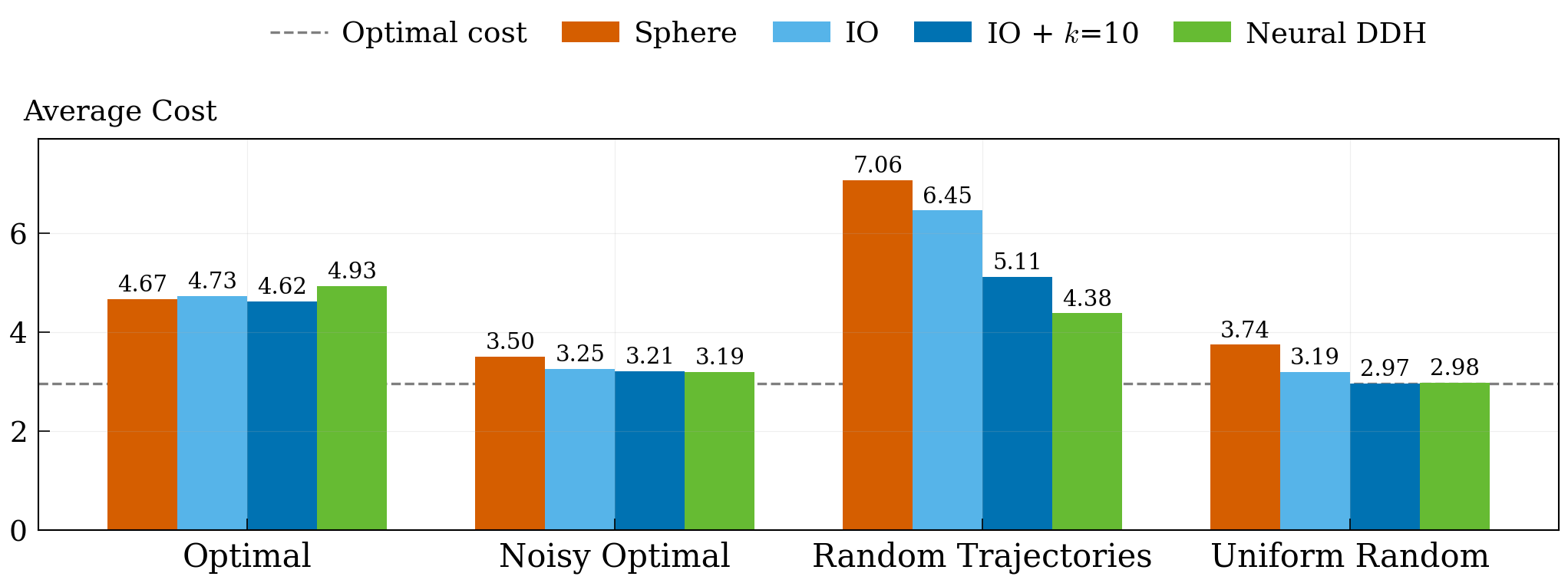}

\caption{\textbf{Each component of the DDH contributes to controller quality.} Average rollout cost over the $1{,}000$ test initial conditions at $n = 500$; dashed line: optimal reference. The rectangular (IO) uncertainty set and tightening each lower the cost on most datasets, and the neural DDH stays close to the grid solution. All variants are in \Cref{tab:lqr-ddh-variants}.}

\label{fig:ddh-variants-comparisons}
\end{figure}

We consider four notions of sensitivity; the first is our running example. Each encodes a different amount of knowledge about $f$ and gives a differently shaped uncertainty set. The uniform constant (i) treats all directions the same and gives a sphere. The input-wise vector (ii) allows a different sensitivity per state coordinate, shrinking the sphere when the state displacement lies in low-sensitivity coordinates. The output-wise vector (iii) bounds each velocity component separately, giving a rectangle. The sensitivity matrix (iv) combines both and is the tightest: entry $L^{\text{io}}_{ij}$ bounds how much state $j$ can change velocity component $i$, so zero entries in the Jacobian become zero-width directions of the set. For the double integrator only one entry is nonzero, which is why the rectangular variants win in \Cref{fig:ddh-variants-comparisons}.

\noindent (i) a uniform Lipschitz constant $\Lx$, satisfying \vspace{-0.25em}
\begin{equation}
    \norm{f(\vect{x}, \vect{u}) - f(\vect{x}', \vect{u})} \le \Lx \norm{\vect{x} - \vect{x}'}, \vspace{-0.25em}
    \label{eq:lipschitz-uniform-appdx}
\end{equation}
\noindent (ii) an input-element-wise constant vector $\Lin$ satisfying
\vspace{-0.5em}
\begin{equation}
    \norm{f(\vect{x}, \vect{u}) - f(\vect{x}', \vect{u})} \le \sum_{j=1}^{\xdim} \Lin_j \lvert x_j - x_j' \rvert 
    \label{eq:lipschitz-input}
\vspace{-1em}
\end{equation}
(iii) an output-element-wise constant vector $\Lout$ satisfying for $i\!=\!1,\!\cdots\!,\!\xdim,$     
\begin{equation}
    \lvert f_i(\vect{x}, \vect{u}) - f_i(\vect{x}', \vect{u}) \rvert \le \Lout_i \norm{\vect{x} - \vect{x}'} 
    \label{eq:lipschitz-output}
\end{equation}
\noindent (iv) an $\xdim \!\times\!\xdim$ sensitivity matrix $\Lio$ satisfying for $i\!=\!1,\!\cdots\!,\!\xdim,$
\begin{equation}
    \lvert f_i(\vect{x}, \vect{u}) - f_i(\vect{x}', \vect{u}) \rvert \le \sum_{j=1}^{\xdim} L_{ij}^{\text{io}} \lvert x_j - x'_j \rvert
    \label{eq:lipschitz-matrix}
\end{equation}
for all $\vect{x}, \vect{x}' \in \mathcal{X}, \vect{u} \in \mathcal{U}.$
We define $\Lu, \Luin, \Luout$, and $\Luio$ analogously with respect to the control.

These lead to uncertainty sets:
\begin{align*}
    &\Lx: \uncertainset^{(i)}(\vect{x}, \vect{u}^{(i)}; \vect{x}^{(i)}, \vect{u}^{(i)}) = \ball\left(\vect{v}^{(i)}, \Lx \norm{\vect{x} - \vect{x}^{(i)}}\right) \\ 
    &\Lin:  \uncertainset^{(i)}(\vect{x}, \vect{u}^{(i)}; \vect{x}^{(i)}, \vect{u}^{(i)}) = \ball\left(\vect{v}^{(i)}, \LinT |\vect{x} - \vect{x}^{(i)}|\right) \\ 
    &\Lout: \uncertainset^{(i)}(\vect{x}, \vect{u}^{(i)}; \vect{x}^{(i)}, \vect{u}^{(i)}) = \rect\left(\vect{v}^{(i)}, 
    \Lout \norm{\vect{x} - \vect{x}^{(i)}}\right) \\
    &\Lio: \uncertainset^{(i)}(\vect{x}, \vect{u}^{(i)}; \vect{x}^{(i)}, \vect{u}^{(i)}) = \rect\left(\vect{v}^{(i)}, \Lio |\vect{x} - \vect{x}^{(i)}|\right)
\end{align*}

Searching over arbitrary controls, as in the \tightenedDDH, adds the corresponding control sensitivity. Writing $f(\vect{x}, \vect{u}) - \vect{v}^{(i)} = \big(f(\vect{x}, \vect{u}) - f(\vect{x}^{(i)}, \vect{u})\big) + \big(f(\vect{x}^{(i)}, \vect{u}) - f(\vect{x}^{(i)}, \vect{u}^{(i)})\big)$ and applying the triangle inequality gives
\begin{align*}
    &\Lx, \Lu: \uncertainset^{(i)}(\vect{x}, \vect{u}; \vect{x}^{(i)}, \vect{u}^{(i)}) = \ball\left(\vect{v}^{(i)}, \Lx \norm{\vect{x} - \vect{x}^{(i)}} + \Lu \norm{\vect{u} - \vect{u}^{(i)}}\right) \\
    &\Lin, \Luin: \uncertainset^{(i)}(\vect{x}, \vect{u}; \vect{x}^{(i)}, \vect{u}^{(i)}) = \ball\left(\vect{v}^{(i)}, \LinT |\vect{x} - \vect{x}^{(i)}| + {\Luin}^\top |\vect{u} - \vect{u}^{(i)}|\right) \\
    &\Lout, \Luout: \uncertainset^{(i)}(\vect{x}, \vect{u}; \vect{x}^{(i)}, \vect{u}^{(i)}) = \rect\left(\vect{v}^{(i)}, \Lout \norm{\vect{x} - \vect{x}^{(i)}} + \Luout \norm{\vect{u} - \vect{u}^{(i)}}\right) \\
    &\Lio, \Luio: \uncertainset^{(i)}(\vect{x}, \vect{u}; \vect{x}^{(i)}, \vect{u}^{(i)}) = \rect\left(\vect{v}^{(i)}, \Lio |\vect{x} - \vect{x}^{(i)}| + \Luio |\vect{u} - \vect{u}^{(i)}|\right)
\end{align*}
Setting $\vect{u} = \vect{u}^{(i)}$ recovers the sets above; the tightened rectangular variants in our experiments use the set in the final row.

\subsection{LQR Baseline Training Details}
\label{appendix:lqr-baselines}
\begin{figure}[ht]
\centering
\includegraphics[width=\linewidth]{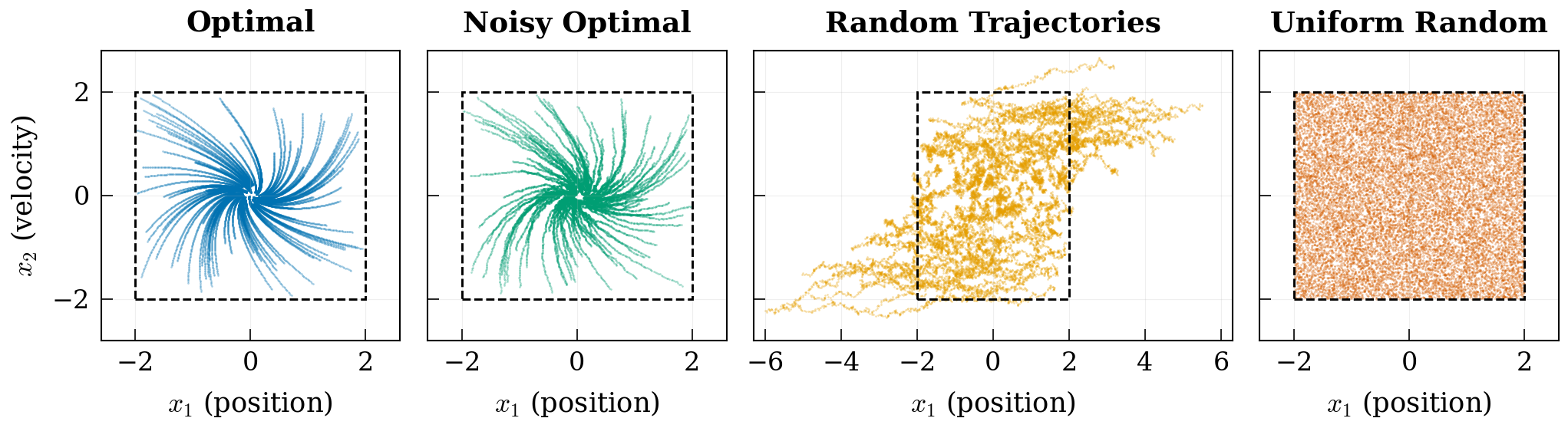}
\caption{\textbf{LQR datasets.} States in the first $20{,}000$ transitions of each dataset; smaller $n$ use a prefix of these. Dashed box: state space $\mathcal{X} = [-2, 2]^2$. \texttt{random trajectories} leaves $\mathcal{X}$ ($26\%$ of its states), so its panel has a wider $x_1$ axis.}
\label{fig:lqr-datasets}
\end{figure}

\xhdr{Datasets} \Cref{fig:lqr-datasets} shows the states each dataset covers. \texttt{noisy optimal} uses
$\vect{\varepsilon} \sim \mathcal{N}(\vect{0}, 0.5^2 I)$. Rewards are the negated
running cost times $\Delta t$, with the negated terminal cost added at the last step of each trajectory; actions are normalized to $[-1,1]$ for training and unnormalized for evaluation. The baselines maximize a discounted return ($\gamma = 0.995$) while evaluation uses the undiscounted finite-horizon cost; this mismatch costs little, since CQL comes within $1\%$ of $J^\star$ on \texttt{optimal}.

\xhdr{Evaluation and epoch selection} Every method is scored by mean
rollout cost from the same $1{,}000$ fixed test initial conditions under the
true dynamics ($T = 2\,$s, $\Delta t = 0.01$). The DDH variants are
deterministic given a dataset. For the offline-RL baselines we hold out $500$ validation initial conditions (disjoint from the test set) and report test cost at the validation-best checkpoint at every dataset size.

\xhdr{CQL} Actor and twin critics are $3$-hidden-layer $\times 256$ ReLU
MLPs. Policy learning rate $10^{-4}$, critic $3 \times 10^{-4}$, batch
size $256$, discount $0.995$, target-network rate $0.005$, deterministic
backup. The entropy coefficient is auto-tuned to target entropy $-2$;
the conservative penalty uses the Lagrangian formulation with target
action gap $5.0$ and $10$ sampled random actions; no behavior-cloning
warmup. $50$ epochs ($50{,}000$ gradient steps), seed $10$.

\xhdr{FQL} We use the official JAX implementation with its default
architecture and optimizer ($4$-hidden-layer $\times 512$ networks,
learning rate $3 \times 10^{-4}$, batch size $256$, $10$ flow steps),
discount $0.995$, $250{,}000$ gradient steps, seed $0$. Following
\citet{park2025flow}, the behavior-cloning coefficient $\alpha$ is tuned
per dataset: we sweep $\alpha \in \{0.001, 0.01, 0.1, 1, 10\}$ and
select on the validation initial conditions, yielding $10$
(\texttt{optimal}), $1$ (\texttt{noisy optimal}), $0.01$
(\texttt{random trajectories}), and $0.001$ (\texttt{uniform random}),
decreasing monotonically with the action quality of the data. FQL trains
for more gradient steps than CQL; both are reported at their
validation-best checkpoint.

\xhdr{IQL} We use the CORL implementation, with its network depth and learning-rate schedule changed to match CQL: $3$-hidden-layer
$\times 256$ networks, expectile $\tau = 0.7$, inverse temperature
$\beta = 3.0$, discount $0.995$, $50$ epochs of $1{,}000$ gradient
steps, with actions normalized as for CQL. As for the other baselines,
we report test cost at the checkpoint with the best cost on the
validation initial conditions.

\xhdr{PPO and SAC} Stable-Baselines3 implementations trained on the true system (forward Euler, $\Delta t = 0.01$, $T = 2\,$s, actions in $[-4, 4]$, discount $0.995$), one seed each. We evaluate the deterministic policy at saved environment-step checkpoints and report the one saved at the $20{,}000$-step milestone ($20{,}000$ steps for SAC; $43{,}008$ for PPO, whose training rounds each checkpoint interval up to whole $2{,}048$-step rollouts) and the final one ($50{,}000$ for SAC, $526{,}336$ for PPO), which is also each run's lowest cost.

\xhdr{Robustness checks} Rerunning CQL (seeds $10$--$12$) and FQL (seeds
$0$--$2$) at three representative cells gives, on
\texttt{random trajectories} at $n = 20{,}000$, $3.34 \pm 0.01$ for CQL
and $4.77 \pm 0.26$ for FQL (mean $\pm$ SEM over seeds); the reported
$13\%$ and $61\%$ gaps are the seed means. CQL is within $1\%$ of $J^\star$ at $n = 500$ on \texttt{optimal} in every seed, and on \texttt{uniform random} at $n = 20{,}000$ CQL and FQL reach $3.16 \pm 0.01$ and $3.13 \pm 0.01$. The
small-$n$ uniform result is likewise not an artifact of the dataset
draw: across five independent re-draws of the \texttt{uniform random} dataset,
grid DDH (IO, $k = 10$) attains $3.03 \pm 0.04$ at $n = 50$ and
$3.00 \pm 0.04$ at $n = 75$ (mean $\pm$ std over draws, vs $J^\star =
2.96$), within $4\%$ of optimal in every draw.

\subsection{DDH Variants on the LQR Problem}
\label{appendix:ddh-variants}
\begin{table}[h]
\centering
\caption{All DDH variants at $n = 500$: average cost over the $1{,}000$ test
initial conditions. Bold = best per column.}
\label{tab:lqr-ddh-variants}
\begin{tabular}{lrrrr}
\toprule
Variant & \texttt{optimal} & \texttt{noisy optimal} & \texttt{random traj.} & \texttt{uniform random} \\
\midrule
Riccati Optimal        & 2.96 & 2.96 & 2.96 & 2.96 \\
Sphere                 & 4.67 & 3.50 & 7.06 & 3.74 \\
Sphere + $k=1$         & 4.67 & 3.49 & 6.15 & 3.45 \\
Sphere + $k=10$        & 4.67 & 3.49 & 6.09 & 3.22 \\
IO                     & 4.73 & 3.25 & 6.45 & 3.19 \\
IO + $k=1$             & 4.62 & 3.22 & 5.29 & 3.05 \\
IO + $k=10$            & \textbf{4.62} & 3.21 & 5.11 & \textbf{2.97} \\
Neural DDH ($k=10$)    & 4.93 & \textbf{3.19} & \textbf{4.38} & 2.98 \\
\bottomrule
\end{tabular}
\end{table}

\Cref{tab:sphere-curve} tests the minimal-information setting, in which the spherical DDH uses only the scalar constants $\Lx = \Lu = 1$ instead of the entrywise structure of $A$ and $B$. On \texttt{random trajectories} it outperforms every offline RL baseline at every tested size from $n = 500$. On \texttt{uniform random} it does so up to $n = 300$ and at $n = 20{,}000$, while FQL is ahead from $n = 400$ and CQL from $n = 1{,}000$, through $n = 5{,}000$. The rectangular set closes most of the remaining gap to $J^\star$, for example $3.11$ vs $5.11$ on \texttt{uniform random} at $n = 50$.

\begin{table}[h]
\centering
\caption{Minimal-information DDH: average cost over the $1{,}000$ test initial conditions. Sphere variants use only scalar Lipschitz constants.}
\label{tab:sphere-curve}
\begin{tabular}{llrrrrrr}
\toprule
Dataset & $n$ & Sphere & Sphere $k{=}10$ & IO $k{=}10$ & CQL & IQL & FQL \\
\midrule
\texttt{random traj.} & $500$    & 7.06 & 6.09 & 5.11 & 17.05 & 13.36 & 18.29 \\
                      & $1{,}000$  & 3.52 & 3.26 & 2.98 & 15.80 & 10.24 & 9.96 \\
                      & $20{,}000$ & 3.11 & 3.03 & 2.96 & 3.35 & 3.74 & 4.76 \\
\texttt{uniform random}      & $50$     & 5.11 & 4.10 & 3.11 & 11.39 & 40.23 & 24.16 \\
                      & $500$    & 3.74 & 3.22 & 2.97 & 7.14 & 14.16 & 3.38 \\
                      & $20{,}000$ & 3.11 & 3.02 & 2.96 & 3.14 & 5.34 & 3.14 \\
\bottomrule
\end{tabular}
\end{table}

\subsection{Sensitivity to Misspecification of the Lipschitz Bound}
\label{appendix:l-misspec}
The DDH assumes a valid sensitivity bound, so we test what happens when it is overestimated. We scale $\Lio$ and $\Luio$ by $2$ and $4$, re-solve the value function (IO, $k=10$, $n=500$), and evaluate on the $1{,}000$ test ICs (\Cref{tab:l-misspec}). Cost rises gradually: by $4$ to $34\%$ at $2\times$ and $10$ to $61\%$ at $4\times$. Even at $4\times$, the DDH still beats correctly configured CQL on the coverage-rich datasets ($3.68$ vs $7.14$ on \texttt{uniform random}, $8.24$ vs $17.05$ on \texttt{random trajectories}). Underestimating the bound breaks the guarantee and is not tested here.

\begin{table}[h]
\centering
\caption{Grid DDH (IO, $k=10$, $n=500$) with inflated Lipschitz bounds: average cost over the $1{,}000$ test ICs.}
\label{tab:l-misspec}
\begin{tabular}{lrrr}
\toprule
Dataset & true $L$ & $L \times 2$ & $L \times 4$ \\
\midrule
\texttt{optimal}        & 4.62 & 5.19 & 5.53 \\
\texttt{noisy optimal}  & 3.21 & 3.35 & 3.53 \\
\texttt{random traj.}   & 5.11 & 6.84 & 8.24 \\
\texttt{uniform random}        & 2.97 & 3.28 & 3.68 \\
\bottomrule
\end{tabular}
\end{table}

\subsection{Lunar Lander Dynamics and Sensitivity Bounds}
\label{appendix:lunar-lander-details}
As described in \Cref{sec:lunar-lander-results}, we base our modeling of the lunar lander dynamics on \citet{inoue2026lunar}.
The lunar lander has six states and three controls. The states consist of the $x$ and $y$ positions and velocities along with the angle and angular velocity.
We will notate the full state as $\vect{z}$, with
\[
    \vect{z} =
    \begin{bmatrix}
        x \\
        v_x \\
        y \\
        v_y \\
        \theta \\
        \omega
    \end{bmatrix}
\]
and control 
\[
    \vect{u} = \begin{bmatrix}
        u_1 \\ u_2 \\ u_3
    \end{bmatrix}
\]
corresponding to the intensity of the main thruster, right thruster, and left thruster from $0$ to $1$.
With a given mass $m$, angular inertia $I$, horizontal spacing of the side thruster from the center of mass $\delta_x$, vertical spacing of the side thruster below the center of mass $\delta_y$, side thruster angle $\gamma$, maximum main thrust $T_p$, maximum side thrust $T_s$, and gravitational constant $g$, we get the following dynamics:
\[
    \dot{\vect{z}} =
    \begin{bmatrix}
        \dot{x} \\
        \dot{v}_x \\
        \dot{y} \\
        \dot{v}_y \\
        \dot{\theta} \\
        \dot{\omega}
    \end{bmatrix}
    =
    \begin{bmatrix}
         v_x \\
         \frac{1}{m} (-u_1 T_p \sin(\theta) - u_2 T_s \cos(\gamma - \theta) + u_3 T_s \cos(\gamma + \theta)) \\
         v_y \\
         \frac{1}{m} (u_1 T_p \cos(\theta) + u_2 T_s \sin(\gamma - \theta) + u_3 T_s \sin(\gamma + \theta)) - g  \\
         \omega \\
         \frac{1}{I}\left(u_2(\delta_x T_s \sin(\gamma) - \delta_y T_s \cos(\gamma)) + u_3(-\delta_x T_s \sin(\gamma) + \delta_y T_s \cos(\gamma))\right)
    \end{bmatrix}
    =
    f(\vect{z}, \vect{u})
\]
This can be equivalently written in a control-affine form
\[
    f(\vect{z}, \vect{u}) = \dot{\vect{z}} = f(\vect{z}) + g(\vect{z}) \vect{u}
\]
where
\begin{align*}\resizebox{\hsize}{!}{$\displaystyle
    f(\vect{z}) = \begin{bmatrix}
        v_x \\
        0 \\
        v_y \\
        -g \\
        \omega \\
        0
    \end{bmatrix},
    \;g(\vect{z}) = \begin{bmatrix}
        0 & 0 & 0 \\
        -\frac{1}{m}T_p \sin(\theta) & -\frac{1}{m} T_s \cos(\gamma - \theta) & \frac{1}{m} T_s \cos(\gamma + \theta) \\
        0 & 0 & 0 \\
        \frac{1}{m} T_p \cos(\theta) & \frac{1}{m} T_s \sin(\gamma - \theta) & \frac{1}{m} T_s \sin(\gamma + \theta) \\
        0 & 0 & 0 \\
        0 & \frac{1}{I}(\delta_x T_s \sin(\gamma) - \delta_y T_s \cos(\gamma)) & \frac{1}{I}(-\delta_x T_s \sin(\gamma) + \delta_y T_s \cos(\gamma))
    \end{bmatrix}$}
\end{align*}
We assume that each control is bounded in $[0, 1]$. For our experiments in this paper we chose 
$m = 1$, $I = 1$, $\delta_x = 1$, $\delta_y = 1$, $\gamma = 0$, $T_p = 9$, $T_s = 3$ and $g = 1.62$.

\begin{figure}[h]
\centering
\begin{tikzpicture}[scale=1.3, >=stealth]
  \draw[dashed, gray] (0,0) -- (0,1.3);
  \begin{scope}[rotate=15]
    \draw[thick, fill=gray!15] (-0.6,-0.4) rectangle (0.6,0.4);
    \fill (0,0) circle (1.5pt);
    \draw[dashed] (0,0) -- (0,1.3);
    \draw[->, thick, blue] (0,-0.4) -- (0,-1.1) node[below] {\small $u_1 T_p$};
    \fill (0.6,-0.3) circle (1pt);
    \fill (-0.6,-0.3) circle (1pt);
    \draw[->, thick, red] (0.6,-0.3) -- ++(-25:0.8) node[right] {\small $u_2 T_s$};
    \draw[dotted] (0.6,-0.3) -- ++(0.9,0);
    \draw[->] (1.2,-0.3) arc (0:-25:0.6) node[midway, right] {\small $\gamma$};
    \draw[->, thick, red] (-0.6,-0.3) -- ++(205:0.8) node[left] {\small $u_3 T_s$};
    \draw[dotted] (-0.6,-0.3) -- ++(-0.9,0);
    \draw[->] (-1.2,-0.3) arc (180:205:0.6) node[midway, left] {\small $\gamma$};
    \draw[<->] (0,0.55) -- (0.6,0.55) node[midway, above] {\small $\delta_x$};
    \draw[<->] (0.45,0) -- (0.45,-0.3) node[midway, left] {\small $\delta_y$};
  \end{scope}
  \draw[->] (0,1.0) arc (90:105:1.0) node[midway, above] {\small $\theta$};
  \draw[->] (-2.2,-1.8) -- (-1.4,-1.8) node[right] {\small $x$};
  \draw[->] (-2.2,-1.8) -- (-2.2,-1.0) node[above] {\small $y$};
\end{tikzpicture}
\caption{Lunar lander at a positive (counterclockwise) angle $\theta$. Arrows show exhaust direction; each thruster pushes the lander opposite to its arrow. The side thrusters sit at $(\pm\delta_x, -\delta_y)$ from the center of mass, with exhaust tilted $\gamma$ below the lander's horizontal axis, so $\gamma > 0$ gives the side thrust an upward component and $\gamma < 0$ a downward one. The drawing uses $\gamma = 25^\circ$ for visibility; our experiments use $\gamma = 0$.}
\label{fig:lander-diagram}
\end{figure}
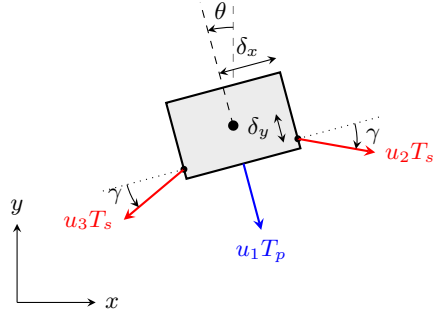

\Cref{fig:lander-diagram} shows the system.

\subsubsection{Sensitivity}
We can find the sensitivity of the dynamics by looking at the Jacobian of the dynamics function $f(\vect{z}, \vect{u})$ with respect to $\vect{z}$ and $\vect{u}$.
\begin{align*}
    &J_{\vect{z}} f(\vect{z}, \vect{u}) = \begin{bmatrix}
            0 & 1 & 0 & 0 & 0 & 0 \\
            0 & 0 & 0 & 0 & -\frac{1}{m}(u_1 T_p \cos(\theta) + u_2 T_s \sin(\gamma - \theta) + u_3 T_s \sin(\gamma + \theta)) & 0 \\
            0 & 0 & 0 & 1 & 0 & 0 \\
            0 & 0 & 0 & 0 & -\frac{1}{m}(u_1 T_p \sin(\theta) + u_2 T_s \cos(\gamma - \theta) - u_3 T_s \cos(\gamma + \theta)) & 0 \\
            0 & 0 & 0 & 0 & 0 & 1 \\
            0 & 0 & 0 & 0 & 0 & 0
    \end{bmatrix} \\
    &J_{\vect{u}} f(\vect{z}, \vect{u}) = g(\vect{z})
\end{align*}
A simple bound for these is:
\begin{align*}
    &|J_{\vect{z}} f(\vect{z}, \vect{u}) | \le \begin{bmatrix}
        0 & 1 & 0 & 0 & 0 & 0 \\
        0 & 0 & 0 & 0 & \frac{1}{m}(T_p + 2T_s) & 0 \\
        0 & 0 & 0 & 1 & 0 & 0 \\
        0 & 0 & 0 & 0 & \frac{1}{m}(T_p + 2T_s) & 0 \\
        0 & 0 & 0 & 0 & 0 & 1 \\
        0 & 0 & 0 & 0 & 0 & 0
    \end{bmatrix}
    \\
    &|J_{\vect{u}} f(\vect{z}, \vect{u}) | \le \begin{bmatrix}
        0 & 0 & 0  \\
        \frac{1}{m} T_p & \frac{1}{m} T_s & \frac{1}{m} T_s \\
        0 & 0 & 0 \\
        \frac{1}{m} T_p & \frac{1}{m} T_s & \frac{1}{m} T_s \\
        0 & 0 & 0 \\
        0 & \left| \frac{1}{I}(\delta_x T_s \sin(\gamma) - \delta_y T_s \cos(\gamma)) \right| & \left| \frac{1}{I}(-\delta_x T_s \sin(\gamma) + \delta_y T_s \cos(\gamma)) \right| 
    \end{bmatrix}
\end{align*}

With a bit more involved trigonometry, we can derive a tighter bound on $|J_{\vect{z}} f(\vect{z}, \vect{u})|$:
\[
|J_{\vect{z}} f(\vect{z}, \vect{u})| \le \begin{bmatrix}
    0 & 1 & 0 & 0 & 0 & 0 \\
    0 & 0 & 0 & 0 & \frac{1}{m}\left(T_p + T_s \max\left(1,\ 2|\sin(\gamma)|\right)\right) & 0 \\
    0 & 0 & 0 & 1 & 0 & 0 \\
    0 & 0 & 0 & 0 & \frac{1}{m}\left(T_p + T_s \max\left(1,\ 2|\sin(\gamma)|\right)\right) & 0 \\
    0 & 0 & 0 & 0 & 0 & 1 \\
    0 & 0 & 0 & 0 & 0 & 0
\end{bmatrix}
\]
which we use in our experiments for $\Lio$.

For the IO method, we can directly use the bounded sensitivity matrices. For the sphere-based method, we instead need the Lipschitz constant, an upper bound on which is given by the maximum singular value of the bounded matrix, $\Lx = \sigma_{\max}(|J_{\vect{z}} f(\vect{z}, \vect{u})|)$ and $\Lu = \sigma_{max}(|J_{\vect{u}} f(\vect{z}, \vect{u})|)$.

\subsection{Lunar Lander Learned Value Functions}
\label{appendix:learned-value-fcns}
\begin{figure}[!t]
\centering
\includegraphics[width=0.9\linewidth]{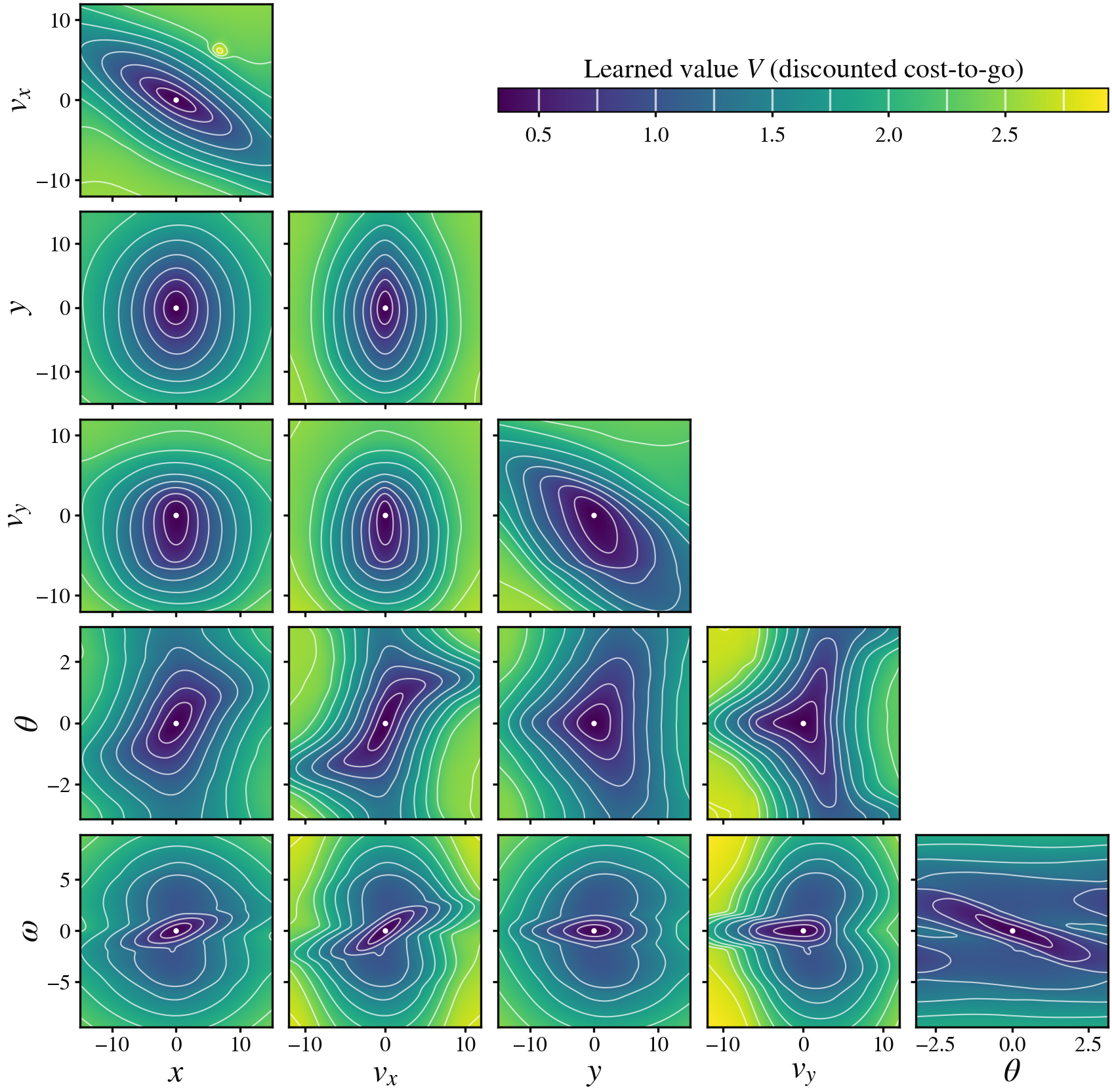}
\caption{\textbf{The DDH learns a qualitatively reasonable value function.} These plots show 2-D slices of the learned value function from the DDH using the known $\Lio$ and known single integrators. We see that they are relatively smooth, displaying few training artifacts.}
\label{fig:value-slices}
\end{figure}

To visualize the value function learned by the DDH, we examine some 2-D slices of  $V_\theta(\vect{x})$. We see qualitatively reasonable slices of the value function in \Cref{fig:value-slices}.

\subsection{Lunar Lander Training Details (DDH and Baselines)}
\label{appendix:lander-training}

For details on the chosen configuration of the lunar lander environment, cost function, and shared parameters between our approaches, see \Cref{tab:hp-setup}.

\xhdr{Neural DDH} We train the neural DDH with \Cref{alg:neural-ddh}. For the first $5{,}000$ gradient steps, the network is trained on the PDE loss $\losspde$ alone ($\lambda_2 = 0$). After that, we add the rollout supervision loss $\lossrollout$ with weight $\lambda_2 = 3$. The supervision set $\datasetSUP$ holds $25{,}000$ rollouts and is regenerated every 240 gradient steps, and the rollout horizon $\tau$ grows linearly from $0.5$ to $4\,$s between steps $5{,}000$ and $15{,}000$. Rollout start states are sampled from $\mathcal{X}_{\text{sup}} = [-7.5, 7.5] \times [-6, 6] \times [-7.5, 7.5] \times [-6, 6] \times [-\pi, \pi] \times [-7.1, 7.1]$ over $(x, v_x, y, v_y, \theta, \omega)$. Note that about $4.8\%$ of the dataset is contained within the supervision region. The lander uses the IO uncertainty set without tightening, so the controller applied at run time is the recorded control selected by the DDH. The uncertainty set for single integrator (``SI'') variants of the DDH incorporates the knowledge that the dynamics of $x, y, \text{ and } \theta$ are given by single integrators of other states, so there is no uncertainty in those dimensions. See \Cref{tab:hp-ddh} for details. 

\begin{table}[h]
\centering\small
\caption{Shared problem parameters.}
\label{tab:hp-setup}
\begin{tabular}{@{}p{0.30\linewidth}p{0.66\linewidth}@{}}
\toprule
Parameter & Value \\
\midrule
Mass, inertia & $m = 1$, $I = 1$ \\
Gravity & $g = 1.62$ \\
Thrust scalings & $T_p = 9$ (main), $T_s = 3$ (side) \\
Lever arms, side-thruster angle & $\delta_x = \delta_y = 1$, $0^\circ$ \\
Hover & $\vect{x}_{\mathrm{eq}} = 0$, $\vect{u}_0 = (mg/T_p, 0, 0) = (0.18, 0, 0)$ \\
$Q$ & $\mathrm{diag}(0.01, 0.01, 0.01, 0.01, 0.1, 0.02)$ \\
$R$ & $0.3\, I_3$ \\
Discount rate $\gamma$ & $0.33\,\mathrm{s}^{-1}$ (per step $e^{-\gamma\Delta t} = 0.993422$) \\
Evaluation horizon $N$ & 757 steps (15.14\,s) \\
State box $\mathcal X$ & $x, y \in [-15,15]$, $v_x,v_y \in [-12,12]$, $\theta \in [-\pi,\pi]$ (periodic), $\omega \in [-3\pi,3\pi]$ \\
Supervision region $\mathcal X_{\mathrm{sup}}$ & $x, y \in [-7.5,7.5]$, $v_x,v_y \in [-6,6]$, $\theta \in [-\pi,\pi]$, $\omega \in [-7.1,7.1]$ \\
Test set & 20{,}000 $\vect{x}_0 \sim \mathrm{Unif}(\mathcal X_{\mathrm{sup}})$ (+20{,}000 in $\mathcal X$) \\
Tuning set & 5{,}000 $\vect{x}_0$ per region, disjoint from test set \\
\bottomrule
\end{tabular}
\end{table}

\begin{table}[h]
\centering\small
\caption{Neural DDH hyperparameters}
\label{tab:hp-ddh}
\begin{tabular}{@{}p{0.30\linewidth}p{0.66\linewidth}@{}}
\toprule
Parameter & Value \\
\midrule
\multicolumn{2}{@{}l}{\emph{Value network}} \\
Architecture & MLP $7 \to 512 \to 512 \to 512 \to 512 \to 1$ (792{,}577 parameters) \\
Activation & Softplus ($\beta = 10$) \\
Input features & state scaled to $[-1,1]$ over $\mathcal X$; $\theta \mapsto (\sin\theta, \cos\theta)$ \\
\midrule
\multicolumn{2}{@{}l}{\emph{Optimization}} \\
Optimizer & Adam, $\beta = (0.9, 0.999)$ \\
Learning rate & $2\times10^{-4}$, constant \\
Training steps & 40{,}000 \\
Pretraining, gradient clipping & none, none \\
PDE points & 25{,}000 per step, resampled, $\mathrm{Unif}(\mathcal X)$ \\
PDE loss weight & 1 \\
Supervised loss weight & 3 \\
\midrule
\multicolumn{2}{@{}l}{\emph{Supervision targets $\hat V$}} \\
Rollouts per refresh & 25{,}000, $\vect{x}_0 \sim \mathrm{Unif}(\mathcal X_{\mathrm{sup}})$ \\
Refresh interval & 240 steps \\
Supervision start & step 5{,}000 \\
Rollout horizon $\tau$ & $\min(4,\ 0.5 + 3.5\times10^{-4}(k - 5000))$\,s \\
Rollout $\Delta t$ & 0.02\,s \\
Horizon bootstrap & $e^{-\gamma \tau} V_\theta(\vect{x}_\tau)$ \\
Exit (leaves $\mathcal X$) & terminate; bootstrap value $1/\gamma = 3.0303$ \\
\midrule
Hardware & 1$\times$ NVIDIA RTX 5090 \\
Wall time & 1\,h\,11\,min \\
\bottomrule
\end{tabular}
\end{table}

\xhdr{Model-based RL} For our model-based RL approach, we have two stages. We first learn a small fully connected neural network to represent an estimate of the dynamics $f$. We use the same dataset of $30$k datapoints that the DDH was trained on. For details on the learning process see \Cref{tab:hp-sysid}. We then apply a certainty-equivalence approach, where we solve a nonlinear MPC assuming our learned model is the true model. Details on the MPC problem (solver, parameters, etc.) can be found in \Cref{tab:hp-mpc}.

\begin{table}[h]
\centering\small
\caption{System identification: learned dynamics model.}
\label{tab:hp-sysid}
\begin{tabular}{@{}p{0.30\linewidth}p{0.66\linewidth}@{}}
\toprule
Parameter & Value \\
\midrule
Architecture & MLP $10 \to 32 \to 32 \to 6$, tanh (1{,}606 parameters) \\
Inputs & $(p_x, v_x, p_y, v_y, \sin\theta, \cos\theta, \omega, u_1, u_2, u_3)$ \\
Loss & MSE, normalized output space \\
Optimizer & Adam \\
Learning rate & $3\times10^{-3}$, cosine to $3\times10^{-6}$ \\
Batch size & 512 \\
Epochs (steps) & 4{,}000 (236{,}000) \\
Seed & 0 \\
Wall time (CPU) & 263\,s \\
\bottomrule
\end{tabular}
\end{table}

\begin{table}[h]
\centering\small
\caption{MPC hyperparameters.}
\label{tab:hp-mpc}
\begin{tabular}{@{}p{0.30\linewidth}p{0.32\linewidth}p{0.32\linewidth}@{}}
\toprule
Parameter & MPC & MPC, 15\,FPS \\
\midrule
Prediction model & Euler step, $\Delta t = 0.02$ \\
Variables & States and controls \\
Horizon $H$ & \multicolumn{2}{p{0.64\linewidth}}{4\,s (200 steps)} \\
Stage cost & \multicolumn{2}{p{0.64\linewidth}}{$\Delta t\, e^{-\gamma j\Delta t} L(\vect{x}_j, \vect{u}_j)$} \\
Terminal cost & \multicolumn{2}{p{0.64\linewidth}}{$w_T L(\vect{x}_H, \vect{u}_0)$, $w_T = \sum_{j\ge H}\Delta t\, e^{-\gamma j\Delta t} = 0.8122$} \\
Constraints & \multicolumn{2}{p{0.64\linewidth}}{$\vect{u} \in [0,1]^3$; no state constraints} \\
Solver & \multicolumn{2}{p{0.64\linewidth}}{IPOPT, MUMPS, exact Hessian, tol $10^{-8}$, CasADi 3.8 (JIT)} \\
Initial guesses ($t = 0$) & \multicolumn{2}{p{0.64\linewidth}}{hover + 5 winding branches $n \in \{-2,\dots,2\}$ (11 solves)} \\
Re-plan interval & 25 steps (0.5\,s) & 3 steps (60\,ms) \\
IPOPT iteration cap & 50 & 6 \\
Warm start & shifted previous plan & shifted primal--dual, $\mu_{\mathrm{init}} = 10^{-4}$ \\
Median re-plan time$^\dagger$ & 360\,ms & 57\,ms \\
\bottomrule
\multicolumn{3}{@{}l}{\footnotesize $^\dagger$ one pinned core, AMD Threadripper PRO 7985WX; $t = 0$ plan excluded.}
\end{tabular}
\end{table}

\xhdr{Model-free RL} We used the original authors' CQL and FQL implementations. For details on their training, see \Cref{tab:hp-cql-fql}. We also used existing implementations of PPO and SAC from the Stable-Baselines3 repository. For details on their training, see \Cref{tab:hp-ppo-sac}. The reset distribution for PPO and SAC was set to match the supervision region for the DDH.

\begin{table}[h]
\centering\small
\caption{PPO and SAC hyperparameters (Stable-Baselines3 2.9.0).}
\label{tab:hp-ppo-sac}
\begin{tabular}{@{}p{0.30\linewidth}p{0.32\linewidth}p{0.32\linewidth}@{}}
\toprule
Parameter & PPO & SAC \\
\midrule
\multicolumn{3}{@{}l}{\emph{Environment}} \\
Observation & 7-D, $\theta \mapsto (\sin\theta,\cos\theta)$ & 7-D, $\theta \mapsto (\sin\theta,\cos\theta)$ \\
Reward Scaling & --- & 10 \\
Reward Normalization & running return std & none \\
Reset distribution & $\mathrm{Unif}(\mathcal X_{\mathrm{sup}})$ & $\mathrm{Unif}(\mathcal X_{\mathrm{sup}})$ \\
Episode length (train) & 152 steps (3.04\,s) & 757 steps (15.14\,s) \\
Termination at $\partial\mathcal X$ & none & none \\
\midrule
\multicolumn{3}{@{}l}{\emph{Networks}} \\
Actor & $7 \to 256 \to 256 \to 3$, tanh & $7 \to 512 \to 512 \to 3$, ReLU \\
Policy distribution & Gaussian, state-independent $\sigma$ & tanh-Gaussian \\
Critic & $V$: $7 \to 256 \to 256 \to 1$, tanh & $2\times Q$: $10 \to 512 \to 512 \to 1$, ReLU \\
\midrule
\multicolumn{3}{@{}l}{\emph{Optimization}} \\
Optimizer & Adam ($\epsilon = 10^{-5}$) & Adam ($\epsilon = 10^{-8}$)\\
Learning rate & $3\times10^{-4}$, constant & $3\times10^{-4}$, constant \\
Rollout steps per env & 1024 & --- \\
Replay buffer & --- & $10^6$ \\
Minibatch size & 1024 & 256 \\
Epochs per update & 10 & --- \\
GAE $\lambda$ & 0.95 & --- \\
Clip range & 0.2 & --- \\
Entropy & coefficient 0.005 & automatic, target $-3$ \\
Value loss coefficient & 0.5 & --- \\
Max gradient norm & 0.5 & --- \\
Target smoothing $\tau$ & --- & 0.005 \\
Learning starts & --- & 10{,}000 \\
\midrule
\multicolumn{3}{@{}l}{\emph{Runtime}} \\
Hardware & CPU (1 thread) & CPU (1 thread) \\
Wall time (full budget) & 4.6\,h & 16.7\,h \\
\bottomrule
\end{tabular}
\end{table}

\begin{table}[h]
\centering\small
\caption{CQL and FQL hyperparameters.}
\label{tab:hp-cql-fql}
\begin{tabular}{@{}p{0.30\linewidth}p{0.32\linewidth}p{0.32\linewidth}@{}}
\toprule
Parameter & CQL & FQL \\
\midrule
Implementation & rlkit CQL (reference) & official FQL (JAX) \\
Variant & CQL($\mathcal H$), SAC backbone & flow BC policy + one-step policy \\
Observation & 6-D, $\theta$ wrapped & 7-D, $\theta \mapsto (\sin\theta,\cos\theta)$ \\
Reward scale & 10 & 1 \\
Discount (per step) & 0.993422 & 0.993422 \\
Actor & $6 \to 256^{\times 3} \to 3$, ReLU, tanh-Gaussian & $7 \to 512^{\times 4} \to 3$, GELU (flow and one-step) \\
Critic & $2\times Q$, $256^{\times 3}$, ReLU & $2\times Q$, $512^{\times 4}$, GELU, LayerNorm \\
Target $Q$ aggregation & min & min \\
Conservative weight $\alpha_{\mathrm{CQL}}$ & 0.03 (fixed, no Lagrange) & --- \\
CQL random actions temp. & 10, 1.0 & --- \\
Max-backup & no & --- \\
BC warm-up steps & 0 & --- \\
Distillation coefficient $\alpha$ & --- & 0.001 \\
Flow steps & --- & 10 \\
Entropy & automatic, target $-3$ & --- \\
Learning rate & policy $10^{-4}$, $Q$ $3\times10^{-4}$ & $3\times10^{-4}$ \\
Batch size & 256 & 256 \\
Target smoothing $\tau$ & 0.005 & 0.005 \\
Gradient steps & 100{,}000 & 125{,}000 \\
Evaluation policy & $\tanh(\mu)$ & one-step policy, sampled noise \\
Hardware & 1$\times$ RTX 5090 & 1$\times$ RTX 5090 \\
Wall time per run & 2.3--2.6\,h & 6--7\,min \\
\bottomrule
\end{tabular}
\end{table}

\end{document}